\documentclass[aps,prd,reprint]{revtex4-2}
\usepackage{amsfonts}
\usepackage{amsmath}
\usepackage{amssymb}
\usepackage{mathrsfs}
\usepackage{graphicx}
\usepackage{subfigure}
\usepackage{epstopdf}
\usepackage{float}
\usepackage{color}
\usepackage{cancel}
\usepackage{ulem}
\newcommand{\adn}[1]{{\color{blue}#1}}

\begin{document}
\title{Method of distinguishing between black holes and wormholes}
\author{Wei Hong$^{1,2}$}
\author{Jun Tao$^{1}$}
\email{taojun@scu.edu.cn, corresponding author}
\author{Tong-Jie Zhang$^{2,3}$}
\affiliation{$^{1}$Center for Theoretical Physics, College of Physics, Sichuan University, Chengdu, 610065, China}
\affiliation{$^{2}$Department of Astronomy, Beijing Normal University, Beijing 100875, China}
\affiliation{$^{3}$Institute for Astronomical Science, Dezhou University, Dezhou 253023, People’s Republic of China}

\begin{abstract}
Beginning with a brief review of the regular space-time with asymptotically Minkowski core, we can consider two copies of the space-time connected through a short-throat wormhole whose radius of mouth is equal to or larger than an extremal regular black hole with asymptotically Minkowski core's event horizon radius. If the wormhole is traversable and smooth, fluxes in these two space-times will interact with and flow into each other. On the cosmological scale, gravity is a candidate for the flux. As the gravitational field changes in one space-time, the behaviours of stars around the wormhole will be affected by the other space-time since we assume there exists enough exotic matter to keep the wormhole open. The changes in a gravitational field can be quantized through the gauge invariant perturbations. The variances in orbits of stars can be reflected by changes in the kinematic shifts of photon frequencies. Then, we use this to distinguish between the black hole and wormhole generated by the same space-time line element, since black hole can not connect two space-times and is unaffected by other space-time.
\end{abstract}
\maketitle

\section{Introduction}
As one of the fundamental theories of modern physics, general relativity predicts many celestial bodies such as black hole, white hole, wormhole and so on. Also, it explains many astronomical phenomena such as Mercury's perihelion precession, gravitational redshift, gravitational drag effect, gravitational wave and so forth. The combination of general relativity and quantum theory, promoting the explorations of connotations of the universe. The observational astronomy and gravitational wave astronomy projects \cite{Akiyama:2019cqa,Akiyama:2019brx,Akiyama:2019sww,Akiyama:2019bqs,Akiyama:2019fyp,Akiyama:2019eap,Barausse:2020rsu} have developed rapidly in recent years, allowing us to find new celestial bodies, such as black holes and wormholes. As one of the celestial bodies predicted by general relativity, wormholes are hypothetical objects which have the feature of connecting two distinct universes or two distinct points of the same universe and were first proposed by Flamm \cite{L.Flamm}. Then we can classify wormholes into non-traversable ones, such as the Euclidean wormholes \cite{Coleman:1989zu,Giddings:1987cg,Coleman:1988tj,Hawking:1989vs}. And the traversable ones, such as Einstein-Rosen bridge \cite{Einstein:1935tc}, Wheeler's form \cite{Wheeler:1957mu}, the MT form proposed in the pioneering work of Morris and Thorne \cite{Morris:1988tu,Morris:1988cz} and Lemos et al studied the MT form with a cosmological constant
\cite{Lemos:2003jb}, the thin shell model first proposed by Visser et al \cite{Visser:1995cc,Hochberg:1998ha,Hochberg:1998ii,Visser:2003yf} and there are some extended work on thin shell model: thin shell wormhole with cosmological constant \cite{Lobo:2003xd}, plane symmetric thin shell wormhole \cite{Lemos:2008aj}, thin shell wormhole in Einstein-Gauss-Bonnet gravity \cite{Richarte:2007zz}, thin shell wormhole in Brans-Dicke gravity \cite{Eiroa:2008hv}. And other types shell wormholes are also interesting such as cylindrical wormholes \cite{Richarte:2013lua}, solitonic shell wormhole \cite{Richarte:2010bd} and so on. If wormholes were real, then the space-time topology of the universe would not be trivial or simply connected \cite{Fuller:1962zza}. So far, we still don't have a definitive astronomical proof that wormholes exist. However, researches in wormholes and its related fields also have an important theoretical value that may change our opinion of the standard inflationary cosmological model. Here are some valuable works to compare wormholes to other celestial bodies: wormholes are distinguished from black holes by Einstein rings generated by gravitational lensing \cite{Tsukamoto:2012xs}, Ellis wormholes are distinguished from other usual massive objects \cite{Tsukamoto:2016zdu,Takahashi:2013jqa,Perlick:2003vg,Nakajima:2012pu,Bhattacharya:2010zzb,Yoo:2013cia,Abe:2010ap,Lukmanova:2016czn,Tsukamoto:2012zz,Gibbons:2011rh} by light path deflection in gravitational lensing under weak field approximation, black holes are distinguished from wormholes by assuming that the active galactic nuclei (AGNs) are wormhole mouths rather than supermassive black holes by hypothesising that wormholes emit gamma rays that produce different spectrum \cite{Piotrovich:2020kae} and utilizing the ringdown signature of gravitational waves to probe the event horizon \cite{Cardoso:2016rao}.

The wonderful universe leaves us with huge amounts of data to achieve the goals we want. As far as the current sky surveys' data, we have not found the existence of wormholes. This fact may give us two interesting perspectives. First is that the scales of our sky surveys may be not large enough. Because if our universe really has a non-trivial topological structure like a wormhole, and its scale is much smaller than our sky survey scale, then light from a long distance will encircle the universe many times before reaching us, and we will see the same group of galaxies in the repeated configuration. But in fact, astronomers tell us that we do not see such repetitions, that is to say, this non-trivial topology's scale is larger than any current sky survey's scale. Secondly, it is not so easy for us to separate the observation data of black holes and wormholes, which may be mixed together. The second perspective is the motivation for our research, which aims to provide a method that can be used to distinguish between wormholes and black holes. We will use the kinematic shifts of photon frequencies emitted by stars \cite{Herrera-Aguilar:2015kea,Becerril:2016qxf,Aschenbach:2004kj,Aschenbach:2004fx,Kraniotis:2007zz,Ghez:2008ms} to distinguish wormholes from black holes.

In this paper, we use a simple model of wormhole space-time: two copies of the regular space-time with asymptotically Minkowski core \cite{Berry:2020tky} smoothly connected through a short-throat wormhole \cite{Dai:2019mse,Simonetti:2020vhw,Dai:2019nph,Dai:2020ffw,Dai:2020rnc} whose radius of mouth is equal to or larger than the event horizon radius $\mathcal{R}$ of an extremal regular black hole with asymptotically Minkowski core. We also assume that there is enough exotic matter near the throat of the wormhole to make it open and stable. At the meantime, we also use the black hole model under the same space-time line element \cite{Simpson:2019mud}. If we do not consider the influence of another space-time connected by the wormhole, the motion of stars in these two models are uniform, and frequency shifts of the photons emitted by them are also identical. Then, the redshift/blueshift data of  photons will not be able to distinguish between wormholes and black holes. But as long as we take into account the other space-time which the wormhole is connected \adn{to}, the gravitational perturbation \cite{Chen:2016plo,Regge:1957td,Bardeen:1980kt,Thorne:1980ru,Martel:2005ir,Wardell:2015ada,Gerlach:1979rw,Zerilli:1971wd,Detweiler:2003ci} from the motion of a massive celestial body in it is transmitted to our space-time through the wormhole and affects the motions of stars and the frequency of photons. Whether the change in photon frequency shifts with perturbation and without perturbation is the key to distinguishing wormholes from black holes. We use ``reception-spacetime" and ``test-spacetime" to label the two copies of space-time. Reception-spacetime refers to the space-time where our observer and detector are located in. The main mission of the detector is to detect the frequency shifts of photons emitted by stars orbiting the wormhole. Test-spacetime refers to the space-time where the sources of the massive objects that produce the gravitational perturbations are located in. We let a massive star make a stable orbital motion relative to the wormhole in this space-time, thereby generating continuous and stable gravitational perturbation.

The rest of this paper is as follows: In the Sec. \ref{sectionII}, we give the space-time line element, build the wormhole model, and calculate the frequency shifts of photons arriving at the detector as the star moves around the wormhole in the reception space-time. In the Sec. \ref{sectionIII}, we use the gauge transformation to analyse the gravitational effect of a massive star moving in a circular orbit relative to the wormhole in the test space-time and how the gravitational effect is transmitted through the wormhole to the reception space-time. In the Sec. \ref{sectionIV}, we choose Perihelion as the starting point of receiving gravitational perturbation from the test-spacetime and then calculate the variation of trajectories of the star and the frequency shifts of photons reaching the detector in the reception space-time. In the Sec. \ref{sectionV}, we make a brief discussion and remark. In the Appendix, we provide details of our numerical implementation.

\section{A Brief Review of The Space-time and Construct a Wormhole}

\label{sectionII}

Before constructing our wormhole model, we make a brief review of the space-time of a regular black hole with asymptotically Minkowski core. The line element of the space-time is given by \cite{Berry:2020tky,Simpson:2019mud}
\begin{equation}
\begin{aligned}
	\mathrm{d} s^{2}=&-\left(1-\frac{2 m e^{-a / r}}{r}\right) \mathrm{d} t^{2}+\left(1-\frac{2 m e^{-a / r}}{r}\right)^{-1} \mathrm{d} r^{2}\\
	&+r^{2}\left(\mathrm{d} \theta^{2}+\sin ^{2} \theta \mathrm{d} \phi^{2}\right),
\end{aligned}
\end{equation}
where the mass $m(r)=me^{-a/r}$ called as Misner-Sharp quasi-local mass. The parameter $a$ should be larger than zero, when $|r| \rightarrow 0$, the mass is being exponentially suppressed, which possesses the asymptotically Minkowski core. Otherwise, if $a < 0$ we have an altogether different scenario where asymptotic behaviour for small $r$ indicates massive exponential increase. And as parameter $a=0$, the mass becomes Schwarzschild black hole mass. In our paper, we focus on the parameter $a>0$. Mostly, the metric is $C^{\infty}$ smooth but not $C^{\omega}$ analytic at coordinate location $r=0$, and this property is very important for us to construct our wormhole model later. The black hole's horizon locates at 
\begin{equation}
	r_{H}=2 m e^{W\left(-\frac{a}{2 m}\right)}=\frac{a}{\left|W\left(-\frac{a}{2 m}\right)\right|},
\end{equation}
where $W(x)$ is the real-valued Lambert $W$ function. The existence and number of the black hole's horizon  is strictly limited by the parameter $a$. For $0<a<2m/e$, one has inner horizon $r_{H-}$ and outer horizon $r_{H+}$
\begin{equation}
r_{H^{+}}=2 m e^{W_{0}\left(-\frac{a}{2 m}\right)}, \quad r_{H^{-}}=2 m e^{W_{-1}\left(-\frac{a}{2 m}\right)},
\end{equation}
and $r_{H^{+}}>a>r_{H^{-}}$. For $a=2m/e$, one can find the two horizons merge at $r_{H^{\pm}}=a$ and this case is what we consider in this paper. And, for $a>2m/e$, the horizon locations are undefined and we shall deal with a horizonless compact object. 

Now, we construct our model of wormhole space-time. The two copies of the regular space-time with asymptotically Minkowski core are smoothly connected through a short-throat of radius $\mathcal{R}=r_{H}$ with the parameter $a=2m/e$ which is also the radius of the wormhole mouth, as we want to distinguish between black hole and wormhole in the same metric. Moreover, we are aware of that the traversable wormholes need ``exotic matter'' to keep it open, which is the violations of the averaged null energy condition \cite{Morris:1988tu,Morris:1988cz,Friedman:1993ty,Hochberg:1998ii,Hochberg:1998ha,Visser:2003yf}.  Using the Einstein field equations, the bulk spacetime has the following stress-energy tensor before perturbation \cite{Berry:2020tky}
\begin{equation}
	\begin{aligned}
		\rho&=-p_{r}=\frac{m a \mathrm{e}^{-a / r}}{4 \pi r^{4}},\\
		p_{t}&=-\frac{m a(a-2 r) \mathrm{e}^{-a / r}}{8 \pi r^{5}}.
	\end{aligned}
\end{equation}
From \cite{Visser:2003yf} we know that $p_r$ is guaranteed to be associated with averaged null energy condition violations, whereas inequalities associated with $p_t$ generically represent normal matter. In this case, we can choose the wormhole's field only deviates from the spacetime in the region from the throat $\mathcal{R}$ out to radius $R'$ as the exotic matter should be restricted to a finite spacetime region for the physically realistic \cite{Visser:2003yf}. Hence, the integral is
\begin{equation}
		\oint p_{r} d V=-2m\left(e^{-a/R'}-e^{-a/\mathcal{R}}\right).
\end{equation} 
As we work in the thin-shell and short-throat wormhole model, we assume that the exotic matter locates in the wormhole throat which implies that $R'\rightarrow \mathcal{R}$. Then the violation $\oint p_{r} d V$ is limited to zero and it does not arise an extremum. Therefore, there always exists enough exotic matter to keep the wormhole open and stable.

\begin{figure}[htbp]
	\centering
	\includegraphics[width=1.0\linewidth]{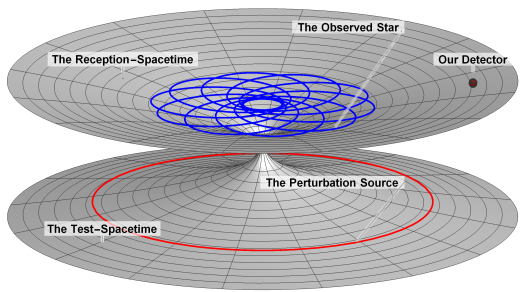}
	\caption{Our model of wormhole space-time. Two copies of the regular space-time with asymptotically Minkowski core smoothly connect through a short-throat wormhole of radius $\mathcal{R}$, which is also the radius of the wormhole mouth. As the figure showed, we choose the upper space-time as the reception-spacetime and we use the blue solid line to show the trajectory of the observed star. Then, we make use of the red solid sphere to exhibit our detector. Corresponding to that, we choose the lower space-time as the test-spacetime and use the red solid line to show the trajectory of the perturbation source.}
	\label{fig:1}
\end{figure}

We choose either of the two space-times to be the place in which we release our detector, and we label the space-time as ``reception-spacetime". Then, the other space-time is labelled as ``test-spacetime". We use the detector to detect the frequency shifts of the photons emitted by stars moving relatively to our wormhole, and we show our wormhole space-time model in FIG. \ref{fig:1}. For simplicity, we make three strong suppositions: (a) The wormhole is located at the focal point of a star's trajectory, regardless of the star's precession. (b) Our detector is located at the equatorial plane of the wormhole which is infinitely far from the wormhole, so that the star we observe can be approximately treated as a particle. (c) At first, in test-spacetime, there is nothing to influence the star's trajectory in reception-spacetime. It means that the star's behaviour is just like it moves relative to a black hole with the event horizon radius $\mathcal{R}$ under the same space-time metric. After a period of time, the gravitational effects in test-spacetime are transmitted to reception-spacetime through the wormhole, causing stars' motions to change. 

\begin{figure}[htbp]
	\centering
	\subfigure[The constant displacement of whole trajectory. Since we place the detector at infinity, the small displacement can be ignored.]{
		\includegraphics[width=1\linewidth]{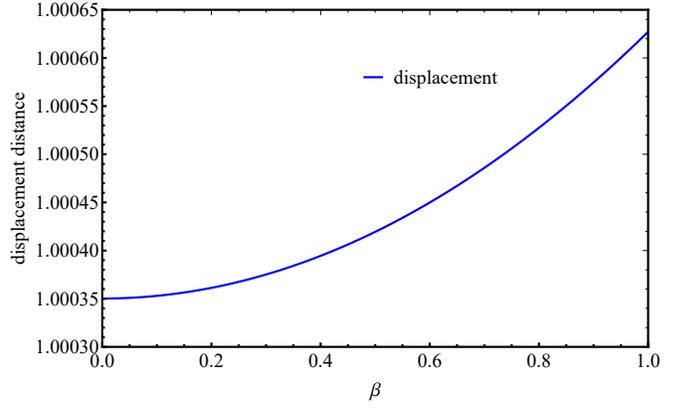}
	}
	\subfigure[The oscillations around zero. From the figure, we can see that the amplitudes of the curves are small and periodic. This is a perfect reflection of a closed circular orbit and four closed elliptical orbits.]{
		\includegraphics[width=1\linewidth]{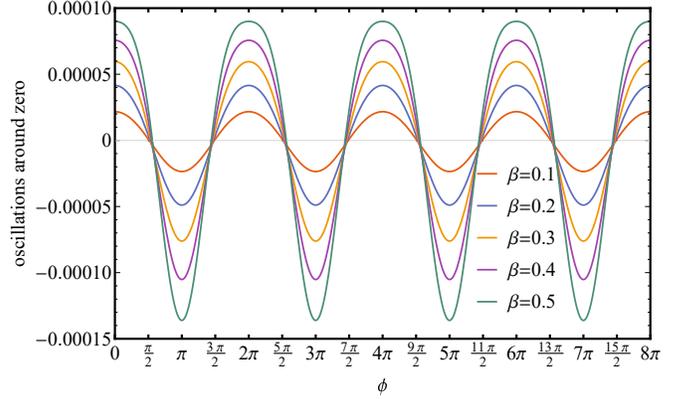}
	}
	\subfigure[The accumulate over successive orbits of the long-axis revolves around the Perihelion. From the figure, we can see that the amplitudes of the curves are not periodic but cumulative. It is useful for us to describe of precession. ]{
		\includegraphics[width=1\linewidth]{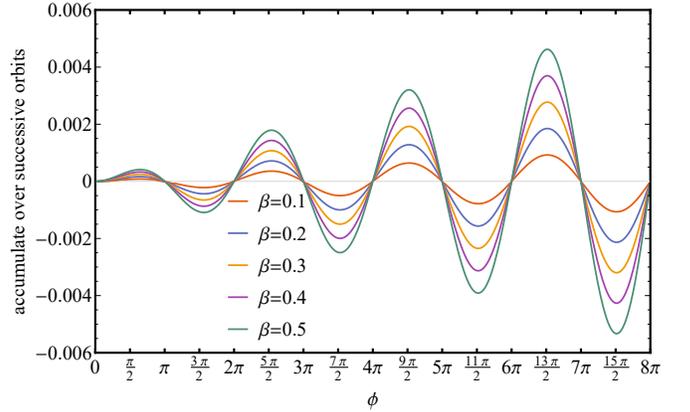}
	}
	\caption{Interpretations of $\tilde{A_1}, \tilde{B_1}$ and $\tilde{C_1}$ in the Eq. (\ref{eq11}). In subfigure (a), we use the blue solid line to show the change in displacement. In subfigure (b) and (c), we use five different colour solid lines to show the five different eccentricities $\beta=$ 0.1, 0.2, 0.3, 0.4 and 0.5 from top to bottom.}
	\label{fig:2}
\end{figure}

 Hence, let us study the trajectories of stars in the regular space-time with asymptotically Minkowski core. To facilitate calculate, we regard stars as particles. The compatibility of the metric means that the inner product of the four-velocity of a particle moving along a geodesic is a constant
\begin{equation}
-g_{\mu\nu}\frac{dx^{\mu}}{d\lambda}\frac{dx^{\nu}}{d\lambda}=1.
\end{equation}
Then, one can obtain
\begin{equation}
	\begin{aligned}
		-\left(1-\frac{2 m e^{-a / r}}{r}\right)\left(\frac{d t}{d \lambda}\right)^{2}&+\left(1-\frac{2 me^{-a / r}}{r}\right)^{-1}\left(\frac{d r}{d \lambda}\right)^{2}\\
		&+r^{2}\left(\frac{d \phi}{d \lambda}\right)^{2}=1,\label{eq5}
	\end{aligned}
\end{equation}
where, we assume that stars are on the $\theta=\frac{\pi}{2}$ plane at the initial moment. Using time-like Killing vector $\xi^{\mu}=(1,0,0,0)$ and space-like Killing vector $\eta^{\mu}=(0,0,0,1)$, one can find two conserved quantities
\begin{equation}
E=\left(1-\frac{2 m e^{-a / r}}{r}\right)\left(\frac{d t}{d \lambda}\right),\quad L=r^{2}\left(\frac{d \phi}{d \lambda}\right).\label{eq6}
\end{equation}
Applying Eq. (\ref{eq6}) into Eq. (\ref{eq5}), we can get a simpler formula
\begin{equation}
	\left(\frac{d r}{d \lambda}\right)^{2}+\left(1-\frac{2 m e^{-a / r}}{r}\right)\left(\frac{L^2}{r^2}+1\right)=E^2.\label{eq7}
\end{equation}
As we know $\frac{d r}{d \lambda}=\frac{d r}{d \phi}\frac{d \phi}{d \lambda}$, we define a new variable quantity $x\equiv\frac{L^2}{mr}$, and then Eq. (\ref{eq7}) becomes to
\begin{equation}
	\begin{aligned}
		&\left(\frac{dx}{d\phi}\right)^2+x^2+2xe^{-\frac{2m^2x}{L^2e}}-\frac{2m^2x^3}{L^2e}e^{-\frac{2m^2x}{L^2e}}\\
		&=\frac{E^2L^2}{m^2}-\left(\frac{L}{m}\right)^2.
	\end{aligned}
\end{equation}
it's worth noting that $e$ is the exponential constant, not the eccentricity $\beta$ of the trajectory. Differentiate the above equation with $\frac{d}{d\phi}$ to get
\begin{equation}
\begin{aligned}
&\frac{d^2 x}{d\phi^2}+x-\left(1-\frac{m^2x}{L^2e}+\frac{3m^2x^2}{L^2}-\frac{2m^4x^3}{L^4e}\right)e^{-\frac{2m^2x}{L^2e}}=0\\
\Rightarrow&\frac{d^2 x}{d\phi^2}-1+x\\
&=-1+\left(1-\frac{m^2x}{L^2e}+\frac{3m^2x^2}{L^2}-\frac{2m^4x^3}{L^4e}\right)e^{-\frac{2m^2x}{L^2e}}.
\end{aligned}
\end{equation}
The left side of the equal sign after the Rightarrow notation is the normalized Binet equation. The right side of the equal sign after the Rightarrow notation can be considered as perturbation correction term, because we use a fact about celestial bodies: $L\gg m$. So, we can write the solution $x$ to a Newtonian solution plus a small deviation as $x=x_0+x_1$ where $x_1$ is the perturbation solution,
\begin{equation}
\begin{aligned}
&\frac{d^2 x_0}{d\phi^2}-1+x_0=0,\\
&\frac{d^2 x_1}{d\phi^2}+x_1\\
&=-1+\left(1-\frac{m^2x_0}{L^2e}+\frac{3m^2x_0^2}{L^2}-\frac{2m^4x_0^3}{L^4e}\right)\left(1-\frac{2m^2x_0}{L^2e}\right).\label{eq10}
\end{aligned}
\end{equation}
Then, we can obtain the solutions of the zeroth-order part $x_{0}$ and the first-order part $x_{1}$
\begin{equation}
\begin{aligned}
x_0=&1+\beta \cos\phi,\\
x_1=&\tilde{A_1}+\tilde{B_1}+\tilde{C_1},\\
\label{eq11}
\end{aligned}
\end{equation}
where $\beta$ is the eccentricity of the orbit. The solution of $x_1$ can be divided into three parts $\tilde{A_1}, \tilde{B_1}$ and $\tilde{C_1}$. The $\tilde{A_1}$ is simply a constant displacement, the $\tilde{B_1}$ is oscillations around zero and the $\tilde{C_1}$ is useful to accumulate over successive orbits. More details for the three parts, see Appendix I. And we show these results separately in FIG. \ref{fig:2}. To plot the FIG. \ref{fig:2}, we fix the Misner-Sharp quasi-local mass $m=1$, the star's angular momentum $L=100a$, the eccentricity of the orbit $\beta\in[0,1)$ and the the rotation angle $\phi\in[0,8\pi]$. The biggest differences between FIG. \ref{fig:2}(b) and FIG. \ref{fig:2}(c) are: (1) The amplitude of the curve in FIG. \ref{fig:2}(c) is not periodic, but it increases over time. (2) The amplitude of the curve in FIG. \ref{fig:2}(c) is much larger than that of the curve in FIG. \ref{fig:2}(b). So $\tilde{C_1}$ is the important part for us to describe the orbit with precession of a star.
Finally, the solution of $x$
\begin{equation}
	\begin{aligned}
		x=&1+\beta \cos\phi+\epsilon\phi\sin\phi\\
		\approx&1+\beta\cos\left(\phi-\epsilon\phi\right),
	\end{aligned}
\end{equation}
where
\begin{equation}
	\begin{aligned}
		\epsilon=&\frac{6 \beta ^3 m^6}{e^2 L^6}+\frac{8 \beta  m^6}{e^2 L^6}-\frac{3 \beta ^3 m^4}{e L^4}+\frac{2 \beta  m^4}{e^2 L^4}\\
		&-\frac{12 \beta  m^4}{e L^4}+\frac{3 \beta  m^2}{L^2}-\frac{3 \beta  m^2}{2 e L^2}\ll 1.
	\end{aligned}
\end{equation}
Hence, we can get 
\begin{equation}
r=\frac{L^2}{m}\frac{1}{1+\beta\cos\left(\phi-\epsilon\phi\right)}=\frac{(1-\beta^2)\alpha}{{1+\beta\cos\left(\phi-\epsilon\phi\right)}}.\label{eq14}
\end{equation}
If $0<\beta<1$, the trajectory is an ellipse. Then, parameter $\alpha$ is semi-major axis. Photons will continue to emit from the star in orbits of Eq. (\ref{eq14}) and will be detected by our detector. We shall characterize the motion of our star by the frequency shifts of photons where the photons are considered to move along null geodesics in the equatorial plane.

\begin{figure*}[htbp]
	\centering
	\subfigure[Kinematic shifts of photons. The photons are emitted by a star in particular trajectories with eccentricity $\beta$ from 0 to 0.5. The star moves away from our detector in the range of rotation angle $\phi$ from 0 to $\pi$. And the star moves toward to our dector in the range of rotation angle $\phi$ from $\pi$ to 2$\pi$.]{
		\includegraphics[width=0.5\linewidth]{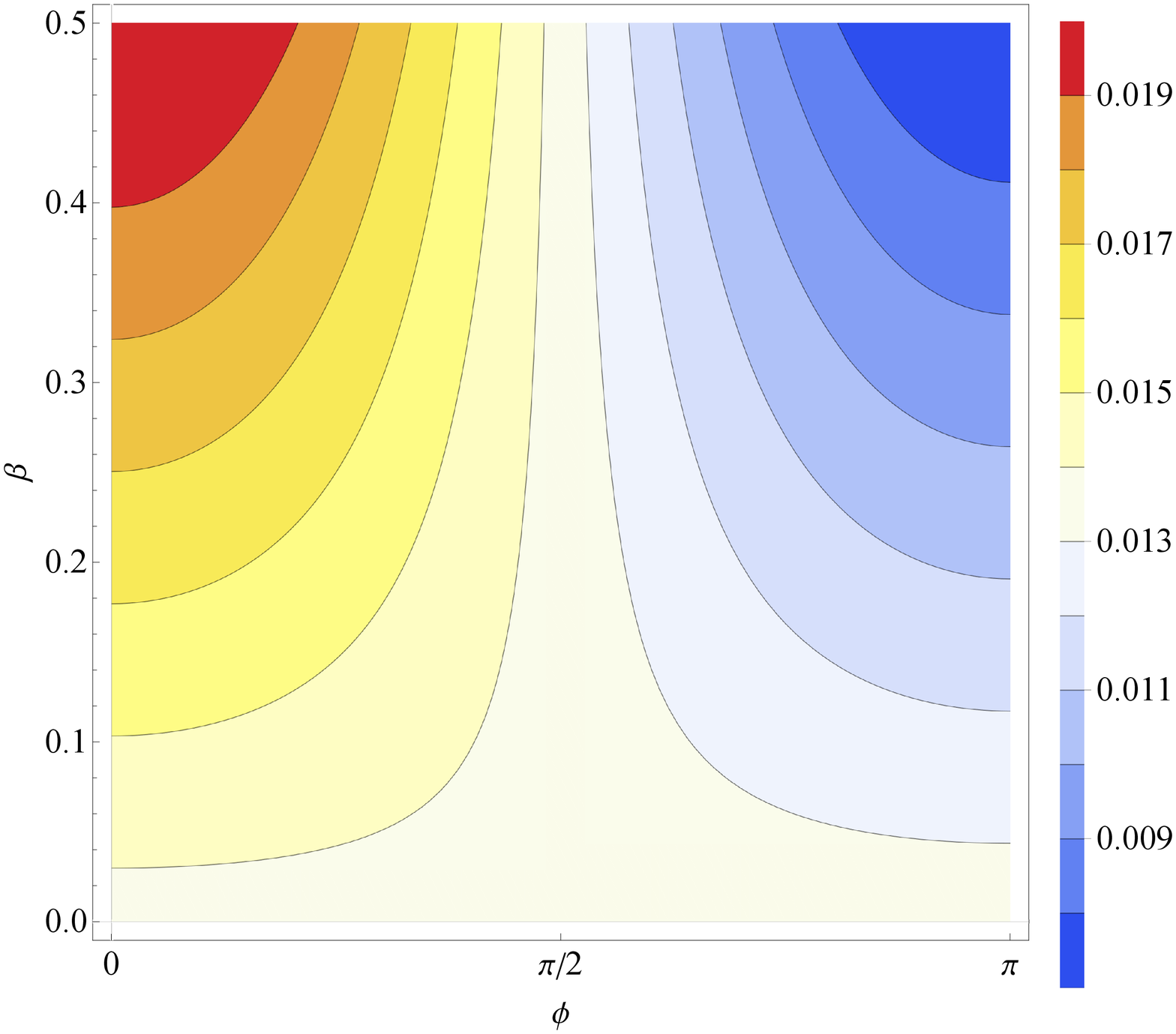} \includegraphics[width=0.5\linewidth]{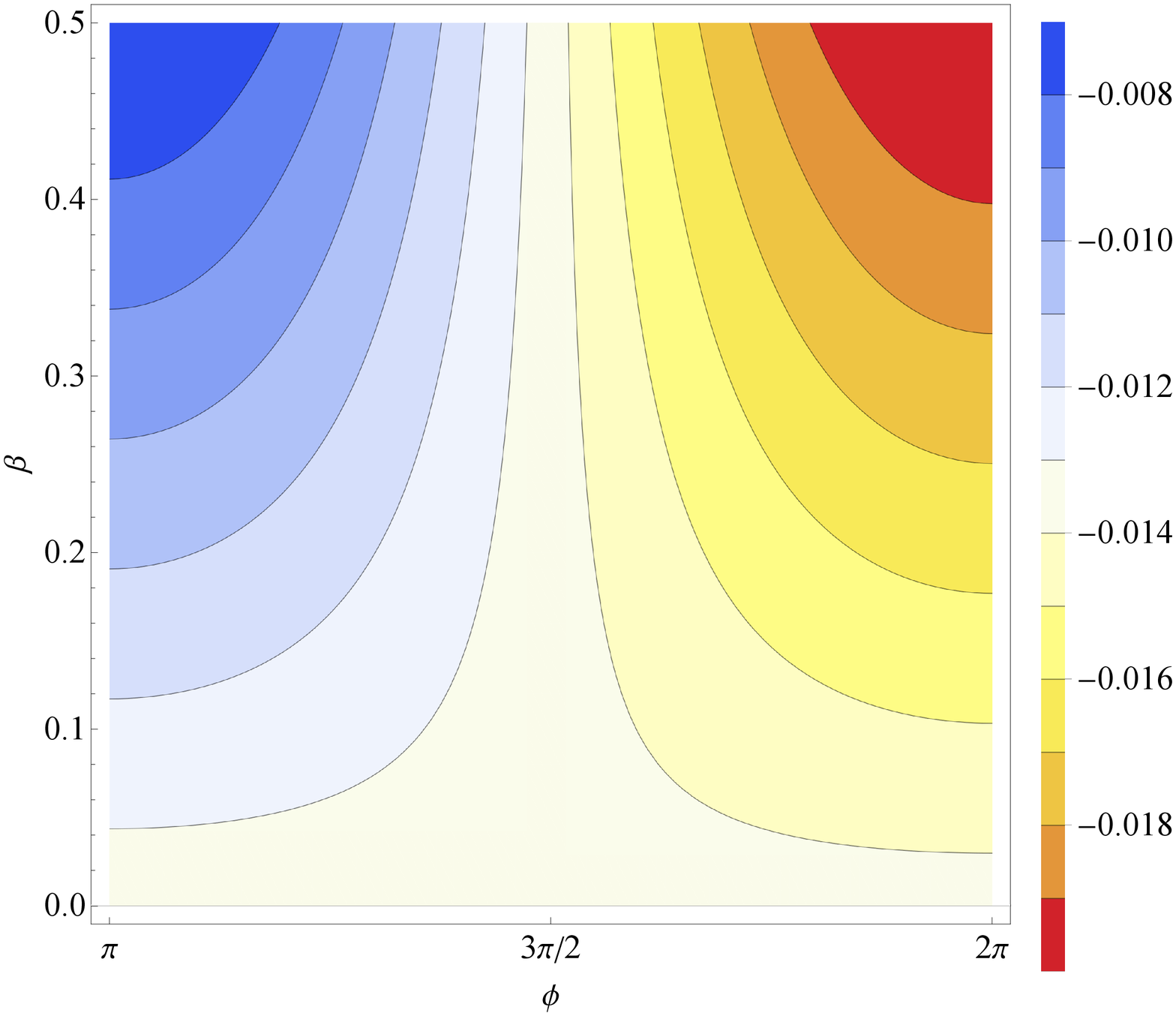}
	}
	\subfigure[Kinematic shifts of photons. The photons are emitted by a star in particular trajectories with eccentricity $\beta$ from 0.5 to 1. The star moves away from our detector in the range of rotation angle $\phi$ from 0 to $\pi$. And the star moves toward to our dector in the range of rotation angle $\phi$ from $\pi$ to 2$\pi$.]{
		\includegraphics[width=0.5\linewidth]{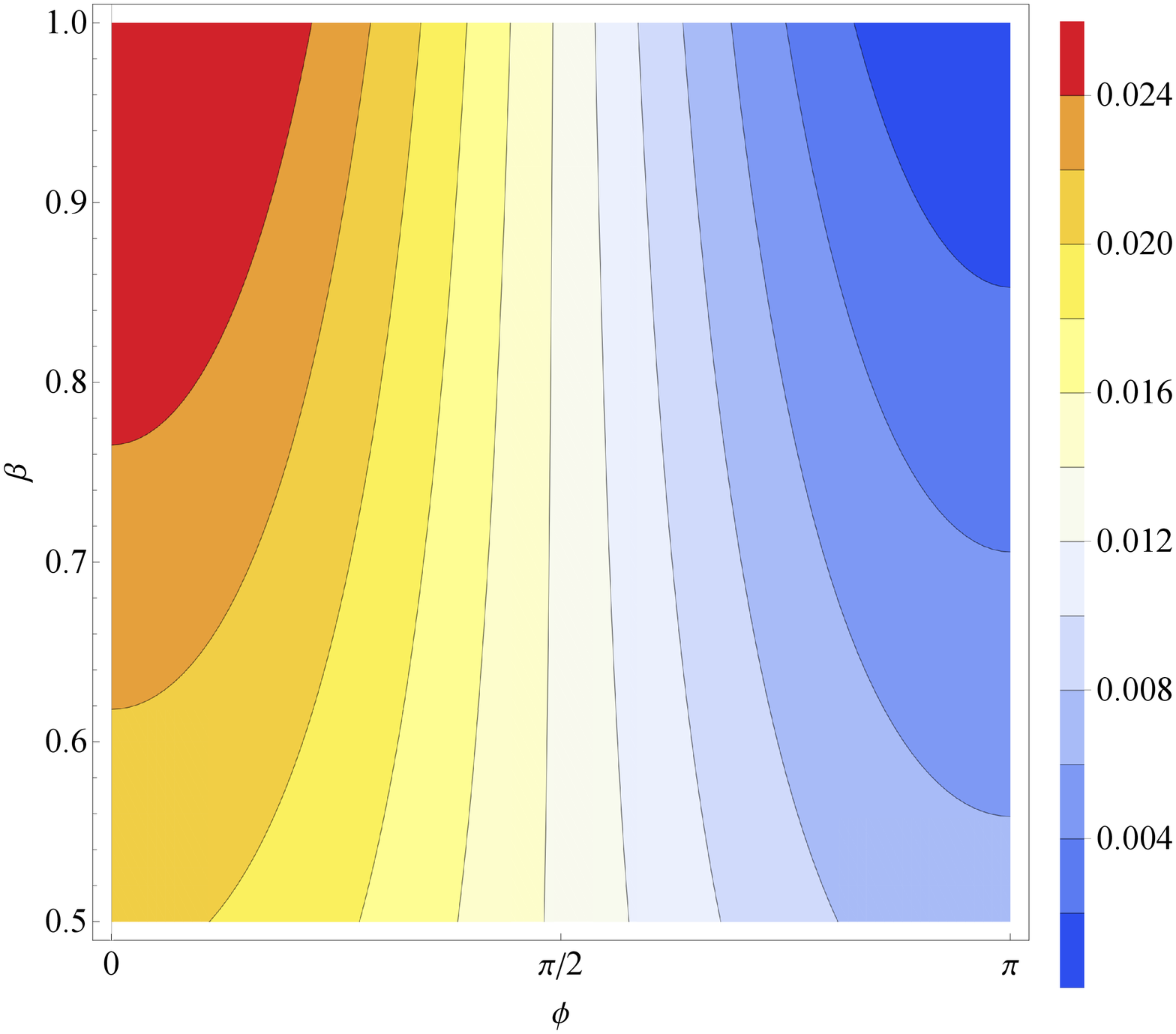} \includegraphics[width=0.5\linewidth]{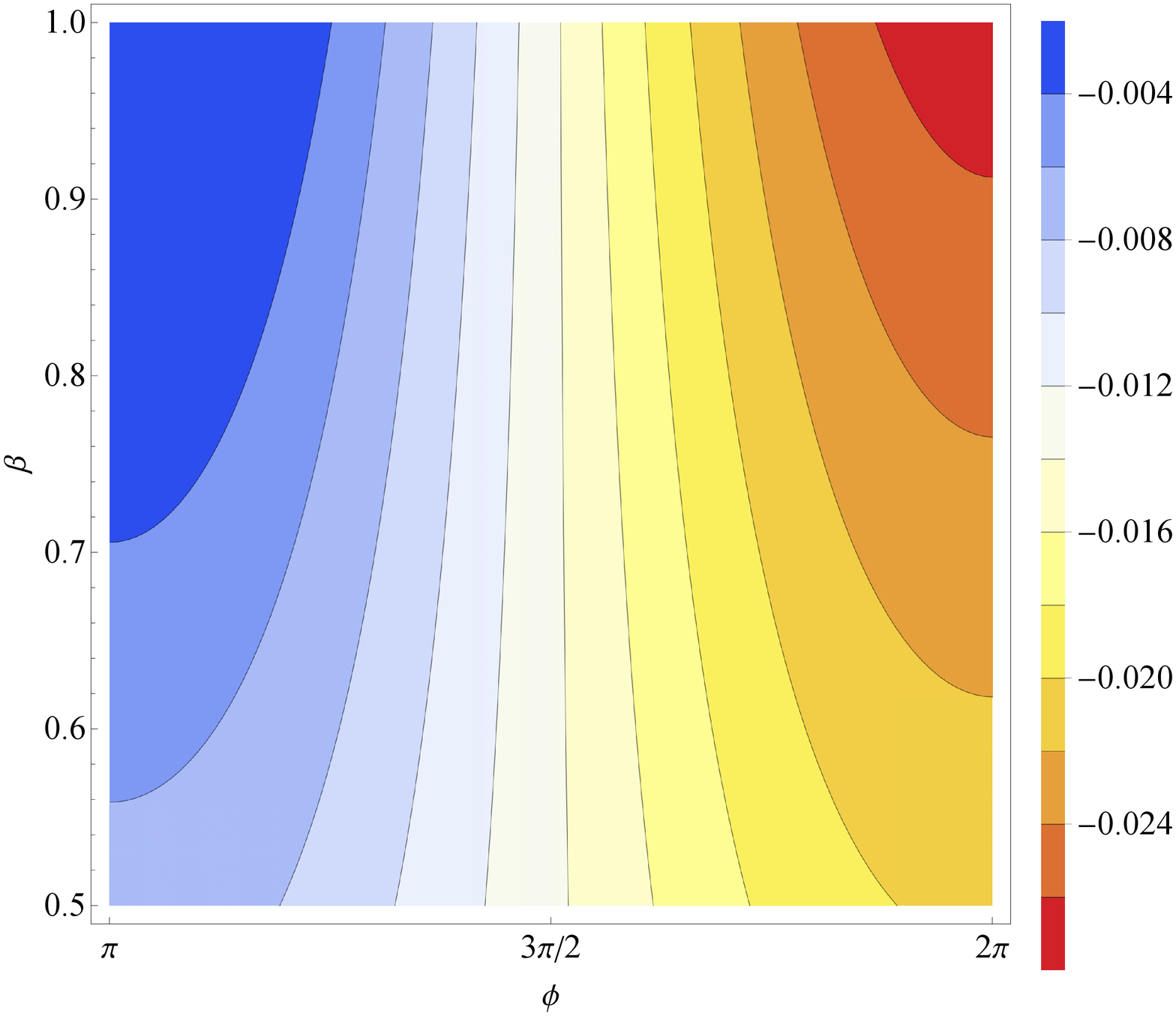}
	}
	\caption{Kinematic shifts of photons in contour plot. The photons are emitted by a star in particular trajectories with eccentricity $\beta$ from 0 to 1. The star moves away from our detector in the range of rotation angle $\phi$ from 0 to $\pi$. And the star moves toward to our detector in the range of rotation angle $\phi$ from $\pi$ to 2$\pi$.}
	\label{fig:3}
\end{figure*}

The frequency shifts $z$ of photons is generally defined as \cite{Herrera-Aguilar:2015kea,Becerril:2016qxf}
\begin{equation}
	1+z=\frac{\nu_{em}}{\nu_{re}}=\frac{-\left. u^{\mu}p_{\mu}\right|_{em}}{-\left. u^{\mu}p_{\mu}\right|_{re}},
\end{equation}
where, $\nu_{em}$ is the frequency emitted by the star at orbit, $p^{\mu}_{em}$ is the photon 4-momentum when it leaves the star, $u^{\mu}_{em}$ is the 4-velocity of the star, $\nu_{re}$ is the frequency received by our detector, $p^{\mu}_{re}$ is the photon 4-momentum received by our detector and $u^{\mu}_{re}$ is the 4-velocity of our detector. Given our previous assumptions, the frequency $\nu_{em}$ and $\nu_{re}$ can be obtained as
\begin{equation}
	\begin{aligned}
		\nu_{em}&=\left.\left(-g_{tt} u^{t} p^{t}-g_{r r} u^{r} p^{r}-g_{\phi\phi} u^{\phi} p^{\phi}\right)\right|_{em},\\
		\nu_{re}&=\left.\left(-g_{tt} u^{t} p^{t}\right)\right|_{re}.
	\end{aligned}
\end{equation}
To simplify the calculation, we introduce three important parameters: energy of photon $E_{\gamma}$, angular momentum of photon $L_{\gamma}$ and the apparent impact parameter $b_{\gamma}$
\begin{equation}
	E_{\gamma}=-g_{tt}p^{t},\quad L_{\gamma}=g_{\phi\phi}p^{\phi}, \quad b_{\gamma}=\frac{L_{\gamma}}{E_{\gamma}}.
\end{equation}
It is worth noting that because photons move along the null geodesics, the energy and angular momentum of photons are preserved in the whole motion. So $b_{\gamma}$ is also invariant throughout the whole null geodesics. Combining them with $p^{\mu}p_{\mu}=0$, one can obtain $(p^{r})^2$, apparent impact parameter $b_{\gamma}$ and frequency shift $z$
\begin{equation}
	\begin{aligned}
		(p^{r})^2&=-\frac{g_{\phi\phi}g^{rr}E_{\gamma}^2+g_{tt}g^{rr}L_{\gamma}^2}{g_{tt}g_{\phi\phi}},\\
		b_{\gamma}&=\pm\sqrt{-\frac{g_{\phi\phi}(r)}{g_{tt}(r)}},\\
		1+z&=\frac{\left.\left(u^{t}-b_{\gamma} u^{\phi}-\frac{1}{E_{\gamma}} g_{r r} u^{r} p^{r}\right)\right|_{em}}{u^t_{re}}.
	\end{aligned}
\end{equation}
Let us consider a practical model of redshift, kinematic redshift $z_{kin}$, as $z_{kin}=z-z_c$. Where $z_c$ corresponds to the frequency shift of a photon emitted by a static particle located at $b=0$,
\begin{equation}
	1+z_c=\frac{u^t_{em}}{u^t_{re}}.
\end{equation}
Therefore, $z_{kin}$ can be written as
\begin{equation}
	\begin{aligned}
		z_{kin}&=(1+z)-(1+z_c)\\
		&=-\frac{\mathscr{U}_{em}+b_{\gamma}u^{\phi}_{em}}{u^t_{re}},\label{eq20}
	\end{aligned}
\end{equation}
where, $\mathscr{U}_{em}$ is a shorthand notation
\begin{equation}
	\mathscr{U}_{em}=\left.\sqrt{-\frac{g_{rr}^2(u^{r})^2}{g_{tt}g_{\phi\phi}}\left(g_{\phi\phi}g^{rr}+g_{tt}g^{rr}b_{\gamma}^2\right)}\right|_{em}.
\end{equation}
We visualize Eq. (\ref{eq20}) as FIG. \ref{fig:3} to present the results and facilitate discussion, . To draw the FIG. \ref{fig:3}, we fix the Misner-Sharp quasi-local mass $m=1$, the star's angular momentum $L=100a$, the eccentricity of the orbit $\beta\in[0,1)$ and the rotation angle $\phi\in[0,2\pi]$. We use different colours to describe the intensity of kinematic redshift or blueshift in photon frequency. The more colour tends to red, the more intense kinematic shifts are. The more colour tends to blue, the less intense kinematic shifts are. Since we start at Perihelion, in the range of rotation angle $\phi\in[0,\pi]$, the star is moving away from our detector, and the frequency of photon has a redshift with a positive value, as shown in the two sub-figures in the left column of FIG. \ref{fig:3}. The more colour tends to red, the more intense redshift is. And in the range of rotation angle $\phi\in[\pi,2\pi]$, the star is moving toward to detector, and the frequency of photon has a blueshift with a negative value, as shown in the two sub-figures in the right column of FIG. \ref{fig:3}. The more colour tends to red, the more intense blueshift is. We can see from the figure that the values of redshift and blueshift are almost symmetrical in absolute value, but not completely symmetrical. Because our star will precess under the action of gravity, its complete period is no longer $2\pi$ but $2\pi(1+\epsilon)$. However, it is a physical fact that the absolute values of redshift and blueshift must be the same and the images must be continuous when the star is at its farthest place from us, even though the sign of apparent impact parameter $b_{\gamma}$ is different. The results shown in FIG. \ref{fig:3} are the same as those shown in the case of a star orbiting a black hole of the same metric. Therefore, to observe a wormhole, one must consider the two space-times connected by the wormhole at the same time. Otherwise, it is impossible to distinguish the wormhole and the black hole under the same metric.

Our next step is to calculate small changes in kinematic shifts by adding the influence of test-spacetime which we did not consider before. There are also celestial bodies in test-spacetime, some of which also move relative to the wormhole and arise additional gravitational effects. The gravitational effects will be transmitted through the wormhole to reception-spacetime, affecting the motion of the star in it. However, the gravitational effects are very weak and need very precise measurement to be observed, so we use perturbation theory to describe one pattern in the gravitational effects.

\section{Gauge Invariant Perturbations of The space-time}

\label{sectionIII}

Gravity can be transmitted from one space-time to another through a traversable wormhole. In our paper, we put a massive star in test-spacetime that also can be viewed as a particle, and the massive star will produce gravitational effects as it moves relative to the wormhole. The gravitational effects can be represented as perturbations on the background metric. In this paper, we use the gauge invariant perturbations \cite{Chen:2016plo,Regge:1957td,Bardeen:1980kt,Thorne:1980ru,Martel:2005ir,Wardell:2015ada,Gerlach:1979rw,Zerilli:1971wd,Detweiler:2003ci} to perturb the test-spacetime. More introduction of the gauge invariant perturbations is shown at the Appendix II. A.

For simplicity, we can write the components of physical metric $g_{ab}$ as an expansion of background metric $g^0_{ab}$ in terms of tensor field. It is useful to choose an orthogonal basis to construct the scalar spherical harmonics and pure-spin vector and tensor harmonics in spherical symmetry of the regular space-time with asymptotically Minkowski core for decomposing tensor fields on the background metric. Hence, we define two unnormalized, constant and orthogonal co-vector fields $v$ and $n$ with components in the coordinates, $v_{a}=(-1,0,0,0), \; n_{a}=(0,1,0,0)$, along with the projection operator onto 2-sphere,
\begin{equation}
	\begin{aligned}
		\sigma_{a b} & \equiv g_{a b}^{0}-\left(1-\frac{2 m e^{-a / r}}{r}\right)^{-1} n_{a} n_{b}+\left(1-\frac{2 m e^{-a / r}}{r}\right) v_{a} v_{b} \\
		&=r^{2} \operatorname{diag}\left(0,0,1, \sin ^{2} \theta\right).
	\end{aligned}
\end{equation}
Then, we can write the metric
\begin{equation}
	g_{a b}=g_{a b}^{0}+h_{a b}.
\end{equation}
Both the physical metric and the background metric are solutions to Einstein Field Equations, and we can expand them in powers of the metric perturbation $h_{ab}$
\begin{equation}
	G_{a b}\left(g^{0}+h\right)=G_{a b}\left(g^{0}\right)-\frac{1}{2} E_{a b}(h)+O\left(h^{2}\right)=8\pi T_{ab},
\end{equation}
where the operator $E_{ab}$ is called linearised Einstein operator
\begin{equation}
	\begin{aligned}
		E_{a b}(h)=&\nabla^{c} \nabla_{c} h_{a b}+\nabla_{a} \nabla_{b} h_{\;c}^{c}-2 \nabla_{(a} \nabla^{c} h_{b) c}\\
		&+2 R_{a\;b}^{\;c\;d} h_{c d}+g_{a b}^{0}\left(\nabla^{c} \nabla^{d} h_{c d}-\nabla^{c} \nabla_{c} h_{\;d}^{d}\right).\label{eq25}
	\end{aligned}
\end{equation}
The notation $\nabla$ is covariant derivative operator and $R_{abcd}$ is the background space-time curvature tensor in our whole paper. We suppose that $E_{ab}$ always satisfies vacuum Einstein Field Equation, so
\begin{equation}
E_{ab}=-16\pi T_{ab}.\label{eq26}
\end{equation}
As long as we solve the Eqs. (\ref{eq25}) and (\ref{eq26}) under certain initial and boundary conditions, we can get the result of metric perturbation we want, but the process is very complicated. Luckily, Detweiler introduced a convenient decomposition of harmonic modes of the metric perturbation
\begin{equation}
	\begin{aligned}
		h_{a b}^{\ell' m'}&=A v_{a} v_{b} Y^{\ell' m'}+2 B v_{(a} Y_{b)}^{E, \ell' m'}+2C v_{(a} Y_{b)}^{B, \ell' m'}\\
		&+2 D v_{(a} Y_{b)}^{R, \ell' m'}+E T_{a b}^{T 0, \ell' m'}+F T_{a b}^{E 2, \ell' m'}+G T_{a b}^{B 2, \ell' m'}\\
		&+2 H T_{a b}^{E 1, \ell' m'}+2 J T_{a b}^{B 1, \ell' m'}+K T_{a b}^{L 0, \ell' m'},\label{eq27}
	\end{aligned}
\end{equation}
where, all parameters \cite{Regge:1957td} from $A$ to $K$ are scalar functions of $(t,r)$, one can find that our perturbation results are not the same as gravitational waves. $Y^{\ell' m'}$ is the scalar spherical harmonic, $Y_{a}^{any', \ell' m'}$ are pure-spin vector harmonics, and $T_{a}^{any', \ell' m'}$ are pure-spin tensor harmonics. They are adapted from Thorne \cite{Thorne:1980ru} with a different normalization by Detweiler \cite{Detweiler:2003ci}
\begin{equation}
	\begin{aligned}
		&Y_{a}^{E, \ell' m'}=r \nabla_{a} Y^{\ell' m'}, \quad Y_{a}^{B, \ell' m'}=r \epsilon_{a b}^{\;\;\; c} n^{b} \nabla_{c} Y^{\ell' m'}, \\
		&Y_{a}^{R, \ell' m'}=n_{a} Y^{\ell' m'},\\
		&T_{a b}^{T 0, \ell' m'}=\sigma_{a b} Y^{\ell' m'}, \quad  T_{a b}^{L 0, \ell' m'}=n_{a} n_{b} Y^{\ell' m'},\\
		&T_{a b}^{B 1, \ell' m'}=r n_{(a} \epsilon_{b)} { }^{d} n^{c} \nabla_{d} Y^{\ell' m'}, \\
		&T_{a b}^{B 2, \ell' m'}=r^{2} \sigma_{(a}^{c} \epsilon_{b) e} d n^{e} \nabla_{c} \nabla_{d} Y^{\ell' m'}, \\
		&T_{a b}^{E 1, \ell' m'}=r n_{(a} \nabla_{b)} Y^{\ell' m'}, \\
		& T_{a b}^{E 2, \ell' m'}=r^{2}\left(\sigma_{a}^{c} \sigma_{b}^{d}-\frac{1}{2} \sigma_{a b} \sigma^{c d}\right) \nabla_{c} \nabla_{d} Y^{\ell' m'}.
	\end{aligned}
\end{equation}
The harmonics are mutually orthogonal
\begin{equation}
	\begin{aligned}
	&\oint Y^{\ell' m'}\left(Y^{\ell'' m''}\right)^{*} \mathrm{d} \Omega=\delta_{\ell' \ell''} \delta_{m' m''},\\
	&\oint Y_{a}^{any', \ell' m'}\left(Y_{any'', \ell'' m''}^{a}\right)^{*} \mathrm{d} \Omega\\&=N(any', r, \ell') \delta_{any' any''} \delta_{\ell' \ell''} \delta_{m' m''},\\ 
	&\oint T_{a b}^{any', \ell' m'}\left(T_{any'', \ell'' m''}^{a b}\right)^{*} \mathrm{d} \Omega\\&=N(any', r, \ell') \delta_{any' any''} \delta_{\ell' \ell''} \delta_{m' m''}.
	\end{aligned}
\end{equation}
The normalization functions $N$ we used are slightly different from \cite{Chen:2016plo}, and when parameter $a\rightarrow0$ the functions reduce to Schwarzschild case. For example,
\begin{equation}
	\begin{aligned}
		&\oint T_{a b}^{L 0, \ell' m'}\left(T_{L 0, \ell'' m''}^{a b}\right)^{*} \mathrm{d} \Omega\\
		&=\left(1-\frac{2 m e^{-a / r}}{r}\right)^{2} \delta_{\ell' \ell''} \delta_{m' m''},\label{eq30}
	\end{aligned}
\end{equation}
and the full expressions of the normalization functions are listed in Appendix II. B.

In this way, we can find the solution of the parameters Eq. (\ref{eq27}) instead  of solving the formula (\ref{eq25}). In formula (\ref{eq27}), one can consider the parameters from $A$ to $K$ as components of metric perturbation $h_{ab}$ projected onto spherical harmonic basis. It is useful to show the $A-K$ components of metric perturbation $h_{ab}$. For example, the $A$ component of the tensor $h_{ab}$ is
\begin{equation}
	h_{A}=\left(1-\frac{2 m e^{-a / r}}{r}\right)^{2} \oint v^{a} v^{b} h_{a b} Y_{\ell' m'}^{*} \mathrm{d} \Omega.
\end{equation}
The parameters $l',l''$ and $m',m''$ in the above equations are related to the trajectory of the object participating in the perturbation. For the full expressions of the $A-K$ components of metric perturbation $h_{ab}$, please see the Appendix II. B. In Sec. II, we discussed the trajectories in the space-time. When Misner-Sharp quasi-local mass is much larger than perturbation object angular momentum $L$ with $Gm\gg L^2$, the object's trajectory can be a circle. This will greatly reduce our calculation, as we can take $l'=l''=0$ and $m'=m''=0$ for circular orbit.

Under circular orbit condition, the perturbation metric can be rewritten as 
\begin{equation}
	h_{a b}=\frac{1}{2 \sqrt{\pi}}\left(A v_{a} v_{b}+2 D v_{(a} n_{b)}+E \sigma_{a b}+Kn_{a} n_{b}\right).\label{eq32}
\end{equation}
Gauge invariants are obtained by gauge transformation of the components of $h_{ab}$ projected onto spherical harmonic basis, for example
\begin{equation}
	\delta=D+\frac{ e^{a/r}r^2}{2 \left(r e^{a/r}-2 m\right)}\frac{\partial ^2}{\partial t^2}E. 
	\label{eq33}
\end{equation}
And the full expressions of gauge invariants are showed in Appendix II. C. Hence, we can obtain the left of Eq. (\ref{eq26}) in spherical harmonic basis with gauge invariants
\begin{equation}
\begin{aligned}
E_{A}=&-\frac{4\left(r-2me^{-a/r}\right)^3}{r^4}\frac{\partial}{\partial r}\psi\\
&-\frac{4\left[r^2-2m(a-r)\right]\left(r-2me^{-a/r}\right)^2}{r^6}\psi,\\
E_{D}=&\frac{4\left(r-2me^{-a/r}\right)}{r^2}\frac{\partial}{\partial t}\psi,\\
E_{E}=&2\frac{\partial^2}{\partial t^2}\psi-\frac{\left[r^2+m(a-r)\right]\left(r-2me^{-a/r}\right)}{r^4}\frac{\partial}{\partial r}\psi\\
&+\frac{4e^{-2a/r}m(a-r)\left(e^{a/r}r-2m\right)}{r}\psi\\
&+2\frac{\partial}{\partial r}o+\frac{2e^{a/r}\left[am+r\left(e^{a/r}r-m\right)\right]}{r\left(e^{a/r}r-2m\right)}o,\\
E_{K}=&\frac{4}{r^2}\psi+\frac{4e^{a/r}}{e^{a/r}r-2m}o.\label{eq34}
\end{aligned}
\end{equation}
Then, we start to work out the right of Eq. (\ref{eq26}). Similarly in Sec. II, we set the massive star moves along the trajectory in the equatorial plane $\theta=\pi/2$. Therefore, we can express the four-velocity in a circular trajectory as $u_{a}=\left(-E_{mp},0,0,L_{mp}\right)$. Where $E_{mp}$ is the massive star energy, and $L_{mp}$ is the massive star angular momentum
\begin{equation}
	\begin{aligned}
		&E_{mp}=\sqrt{\frac{e^{-a/R} \left(R e^{a/R}-2 m\right)^2}{R\left(Re^{a/R}-3 m\right)+a m}},\\
		&L_{mp}=\sqrt{\frac{m R^3-a m R^2}{R\left(Re^{a/R}-3 m\right)+a m}},
	\end{aligned}
\end{equation}
where $R$ is the radius of circular trajectory. So we can write the stress-energy tensor of the massive particle as \cite{Poisson:2011nh}
\begin{equation}
T_{a b}=M_{mp} \int \frac{u_{a} u_{b}}{\sqrt{-g^{0}}} \delta^{(4)}[\Delta s-s(\tau)] \mathrm{d} \tau.\label{eq36}
\end{equation}
Also, we can obtain the linearised Einstein operator spherical harmonic projection components
\begin{equation}
	\begin{aligned}
		E_{A}^{\prime}&=-16 \pi\left(1-\frac{2 me^{-a/R}}{R}\right) \frac{M_{mp} E_{mp}}{R^{2}} \delta(r-R) Y_{00}^{*}\left(\theta, \phi\right),\\
		E_{D}^{\prime}&=0,\\
		E_{E}^{\prime}&=-8 \pi\left(1-\frac{2 me^{-a/R}}{R}\right) \frac{M_{mp} L_{mp}}{R^{4}} \delta(r-R) Y_{00}^{*}\left(\theta, \phi\right),\\
		E_{K}^{\prime}&=0,\label{eq37}
	\end{aligned}
\end{equation}
where $\prime$ is not a derivative notation. Next, we combine Eq. (\ref{eq34}) and Eq. (\ref{eq37}) to solve the gauge invariants $\psi$ and $o$
\begin{equation}
\begin{aligned}
	&\frac{\partial}{\partial r}\psi+\frac{\left[r^2-2m\left(a-r\right)\right]}{r^2\left(r-2me^{-a/r}\right)}\psi=-\frac{r^4}{4\left(r-2me^{-a/r}\right)^3}E_{A}^{\prime},\\
	&o=\frac{e^{a/r}r-2m}{4e^{a/r}}\left(E_{K}^{\prime}-\frac{4}{r^2}\psi\right)=-\frac{e^{a/r}r-2m}{e^{a/r}r^2}\psi,\\
	&\frac{\partial}{\partial t}\psi=\frac{r^2}{4 \left(r-2 m e^{-a/r}\right)}E_{D}^{\prime}=0,\\
	&E_{E}^{\prime}=2\frac{\partial^2}{\partial t^2}\psi-\frac{\left[r^2+m(a-r)\right]\left(r-2me^{-a/r}\right)}{r^4}\frac{\partial}{\partial r}\psi\\
	&\quad\quad+\frac{4e^{-2a/r}m(a-r)\left(e^{a/r}r-2m\right)}{r}\psi.\label{eq38}
\end{aligned}
\end{equation}
Considering the reality in astrophysics, we can suppose that the perturbation vanishes within the trajectory and exponential term $e^{-a/r}$ can expand in powers as $e^{-a/r}=1-a/r+O(r^2)$. Therefore, one can get the $\psi$ and $o$
\begin{equation}
\begin{aligned}
\psi=&2 \sqrt{\pi }\frac{r \eta(r)}{2 a m+r (r-2 m)}\\
&\times\left\lbrace \frac{M_{mp}E_{mp}R^2}{\left[2 a m+R (R-2 m)\right]\eta(R)}\Theta\left(r-R\right)\right\rbrace ,\\
o=&-2 \sqrt{\pi }\frac{\left(a+r-2 m\right)\eta(r)}{(a+r) \left[2 a m+r (r-2 m)\right]}\\
&\times\left\lbrace \frac{M_{mp}E_{mp}R^2}{\left[2 a m+R (R-2 m)\right]\eta(R)}\Theta\left(r-R\right)\right\rbrace,\label{eq39}
\end{aligned}
\end{equation}
where, the expression in curly braces is a constant that depends only on the radius of trajectory where the perturbation source is placed, $\Theta\left(r-R\right)$ is unit step function, $\eta(r)$ and $\eta(R)$ are two useful functions, which can simplify our operations
\begin{equation}
\begin{aligned}
\eta(r)&=\text{exp}\left[-\frac{2 \sqrt{m} }{\sqrt{2 a-m}}\arctan\left(\frac{r-m}{\sqrt{m} \sqrt{2 a-m}}\right)\right], \\
\eta(R)&=\text{exp}\left[-\frac{2 \sqrt{m} }{\sqrt{2 a-m}}\arctan\left(\frac{R-m}{\sqrt{m} \sqrt{2 a-m}}\right)\right].\label{eq40}
\end{aligned}
\end{equation}
One can easily find out from Eq. (\ref{eq40}) 
\begin{equation}
	\frac{\partial}{\partial r}\eta(r)=-\frac{2m}{2am+r(r-2m)}\eta(r).
\end{equation}
Then, we substitute Eqs. (\ref{eq39}) and (\ref{eq40}) into Eq. (\ref{eq33}) to find parameters $A$, and $K$
\begin{equation}
	\begin{aligned}
		A(t,r)=&-4 \sqrt{\pi}\frac{\eta(r)}{r}\kappa(R)T(t)+O\left(\kappa(R)^2\right),\\
		K(r,t)=&4 \sqrt{\pi }\frac{r \eta(r)}{2 a m+r (r-2 m)}\kappa(R)T(t)\\
	            =&4 \sqrt{\pi }\frac{\eta(r)}{r-2me^{-a/r}}\kappa(R)T(t)+O(r^2),\\
		\kappa(R)=&\frac{M_{mp}E_{mp}R^2}{\left[2 a m+R (R-2 m)\right]\eta(R)}\Theta\left(r-R\right),
	\end{aligned}
\end{equation}
where, $T(t)$ is function of time. For circular trajectory with fixed constant radius $R$, the time function can be separated out which does not effect the metric perturbation. $\kappa(R)$ is a constant function of radius $R$ which only depends on the perturbation source. In addition, the parameter $D(t,r)$ and $E(t,r)$ will vanish \cite{Zerilli:1971wd}, as we choose the massive star moving in circular trajectory. Finally, we can obtain the physical metric $g_{ab}$
\begin{equation}
	\begin{aligned}
		g_{tt}&=g^{0}_{tt}+h_{tt}=-\left[1-\frac{2 \left(m e^{-a / r}+\eta(r)\kappa(R)\right)}{r}\right],\\
		g_{rr}&=g^{0}_{rr}+h_{rr}\\
		&=\frac{1}{1-2 \left[m e^{-a / r}+\eta(r)\kappa(R)\right]/r}+O\left(\kappa(R)^2\right),\\
		g_{\theta\theta}&=r^2, \quad g_{\phi\phi}=r^2\sin^2\theta.\label{eq43}
	\end{aligned}
\end{equation}

\begin{figure*}[htbp]
	\centering
	\subfigure[The time-time components of metric in region $r$ from $a$ to 10$a$ and the unit length of the abscissa axis in the figure is $a$. In the region closer to the wormhole, the intensity of the gravitational perturbation is higher.]{
		\includegraphics[width=0.48\linewidth]{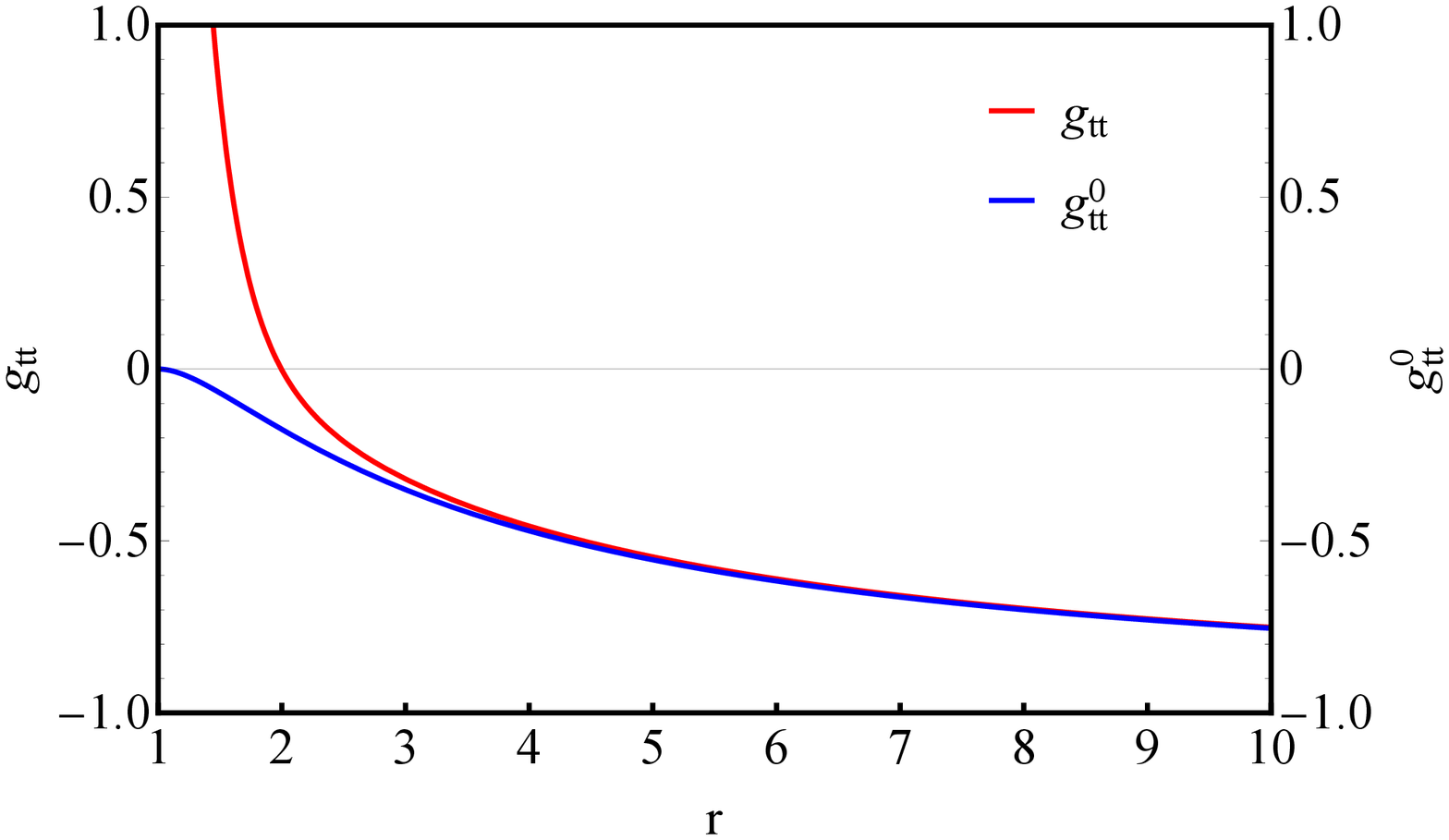}
	}
	\subfigure[The radial-radial components of metric in region $r$ from $a$ to 10$a$ and the unit length of the abscissa axis in the figure is $a$. The trends of the two radial-radial components are almost the same when they are far from the wormhole. However, near the wormhole, gravitational perturbation causes the radial-radial component to branch out into new branches.]{
		\includegraphics[width=0.48\linewidth]{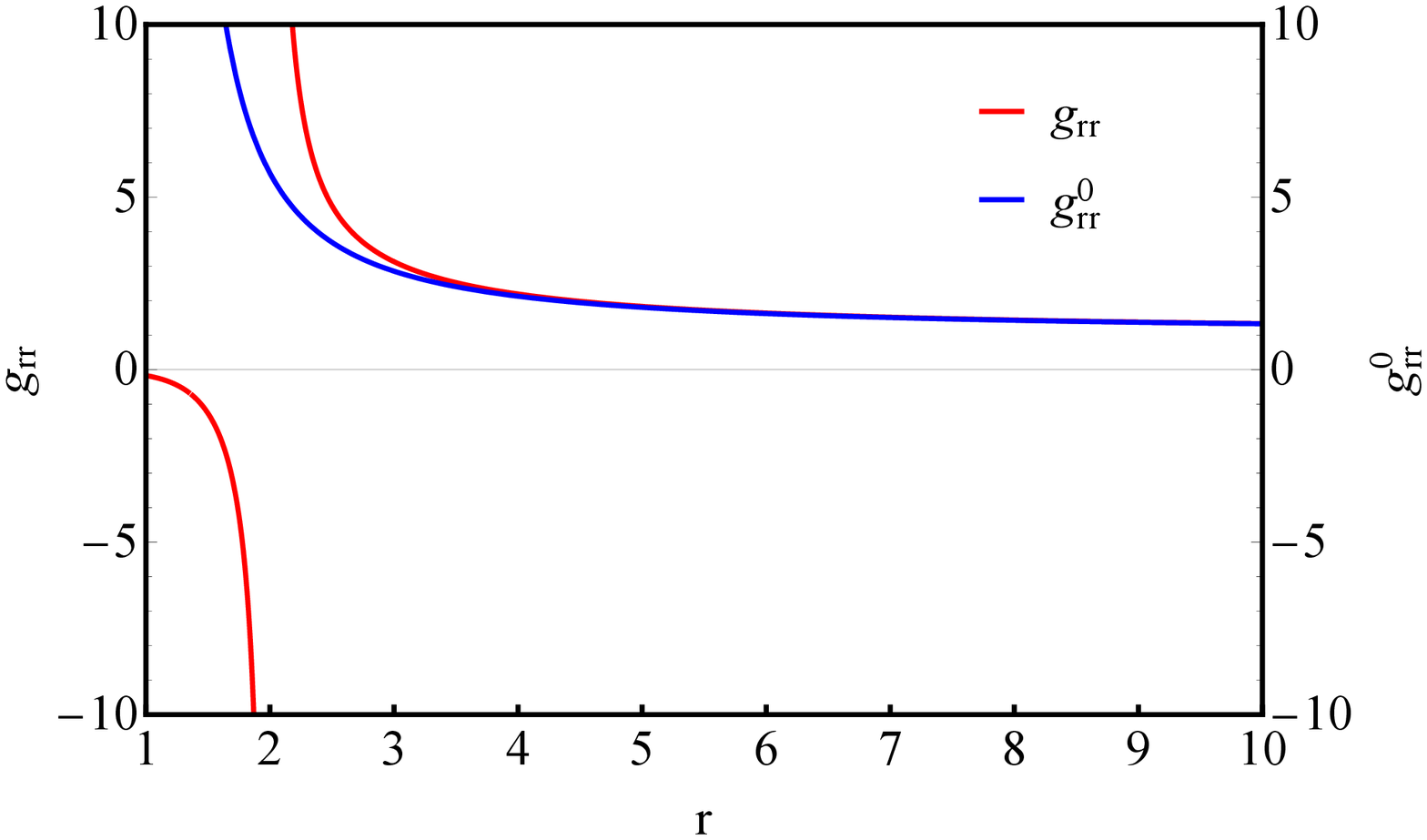}
	}
	\caption{The components of metric with perturbation and without perturbation in region $r\in[a,10a]$ in reception-spacetime. We use the red solid line to signify the components of metric with perturbation and we use the blue solid line to signify the components of metric without perturbation.}
	\label{fig:4}
\end{figure*}

The effect of a massive object moving in a circular orbit on the test-spacetime can be described as Eq. (\ref{eq43}). It can be transmitted through the wormhole. As we mentioned in Sec. II, the wormhole metric is $C^{\infty}$ smooth and two copied of space-time smoothly are connected through a short-throat. Therefore, the gravitational effect satisfies a continuity condition \cite{Dai:2019mse,Simonetti:2020vhw} between test-spacetime effect $h_{ab}$ and reception-spacetime effect $h^{\prime}_{ab}$
\begin{equation}
	h_{ab}(\mathcal{R})=h^{\prime}_{ab}(\mathcal{R}), \quad \left.\frac{\partial h_{ab}}{\partial r}\right|_{r=\mathcal{R}}=\left.\frac{\partial h^{\prime}_{ab}}{\partial r}\right|_{r=\mathcal{R}}.\label{eq44}
\end{equation}
Here, we use a trick that $h_{ab}=h^{\prime}_{ab}$ which satisfies the condition (\ref{eq44}). We visualize metrics in reception-spacetime as  FIG. \ref{fig:4}. We fix these parameters, Misner-Sharp quasi-local mass $m=1.0$, and the perturbation source star mass $M_{mp}=0.01m$, the perturbation source star orbit height $R=30r_{H}$. Where $r_{H}=a=2m/e$ is the radius of wormhole as well as the radius of black hole horizon. The background metric is showed with red solid line and the perturbation source metric is showed with blue solid line. One can find from FIG. \ref{fig:4} that the farther away from the wormhole, the smaller the influence of gravitational perturbation on time and space can be. We will show in the next section how the perturbation affects the orbits of stars in reception-spacetime with the same fixed parameters, and it will also be a method to distinguish wormholes from black holes. 

It is similar to the discussion in the Sec. \ref{sectionII}, we need to see if there exists enough exotic matter to keep the wormhole open. Similarly, the bulk spacetime stress-energy tensor after perturbation can be written as
\begin{equation}
	\begin{aligned}
		\tilde{\rho}&=-\tilde{p_{r}}=\frac{1}{8\pi r^2}\left[\frac{2ame^{-a/r}}{r^2}-\frac{2m\kappa(R)\eta(r)}{2am+r(r-2m)}\right],\\
		\tilde{p_{t}} &=-\frac{m a(a-2 r) \mathrm{e}^{-a / r}}{8 \pi r^{5}}+\frac{m \kappa (R) \eta (r)}{2\pi\left[2 a m+r (r-2 m)\right]^2}.
	\end{aligned}
\end{equation}
In this case, we can choose the wormhole's field only deviates from the spacetime in the region from the throat out to radius $R''$. Hence, the final volume integral is
\begin{equation}
	\begin{aligned}
		\oint p_{r} d V=2m\left(e^{-a/\mathcal{R}}-e^{-a/R''}\right)+\kappa(R)\left(\eta(\mathcal{R})-\eta(R'')\right).
	\end{aligned}
\end{equation} 
As we work in the thin-shell and short-throat wormhole model, we assume that the exotic matter locates in the wormhole throat which implies that $R''\rightarrow \mathcal{R}$. Then the violation $\oint p_{r} d V$ is limit to zero and it does not arise an extremum. Therefore, there also exists the enough exotic matter to keep the wormhole open and stable.

\section{Photon Frequency Redshift and Blueshift of Stars under Wormhole Background}

\label{sectionIV}

In Sec. II, we split the two space-time connected by wormhole and only studied the motions of stars in reception-spacetime without test-spacetime's effects at first. In such a case, it is impossible to distinguish whether the celestial body is moving around a black hole or wormhole. As a consequence, we will consider these two space-times in this section simultaneously for simulating the part of observation phenomena of wormhole. As we hypothesized earlier, the throat of wormhole has enough exotic matter to keep the wormhole from closing due to gravitational perturbations. In this way, perturbations generated in test-spacetime can be transmitted to reception-spacetime and affect the motions of celestial bodies in it. And, small changes in celestial bodies' behaviours are a one of manifestation of wormhole.

We can write the equation of motion like Eq. (\ref{eq7}) in reception-spacetime with perturbation which is generated in test-spacetime by the motion of a massive object
\begin{equation}
	\left(\frac{d r}{d \phi}\right)^{2}+\left(\frac{r^4}{L^2}+r^2\right)\frac{1}{g^{\prime}_{rr}}=-\frac{E^2r^4}{L^2}\frac{1}{g^{\prime}_{rr}g^{\prime}_{tt}}=\frac{E^2r^4}{L^2},\label{eq45}
\end{equation}
where, $g^{\prime}_{\mu\nu}$ is
\begin{equation}
	g^{\prime}_{\mu\nu}=\left(\begin{array}{cccc}
		g^{\prime0}_{tt}+h^{\prime}_{tt} & 0 & 0 & 0 \\
		0 & g^{\prime0}_{rr}+h^{\prime}_{rr} & 0 & 0 \\
		0 & 0 & r^2 & 0 \\
		0 & 0 & 0 &r^2\sin^2\theta
	\end{array}\right).
\end{equation}
It is heavy workload to find the analytic expression for Eq. (\ref{eq45}) directly. Therefore, we need to study the changes of the parameters in the equation to pave the way for our subsequent approximation processing. We express the change as
\begin{equation}
		\Delta_{rr^\prime}=\left(\frac{1}{g^{\prime0}_{rr}}-\frac{1}{g^{\prime}_{rr}}\right)/\frac{1}{g^{\prime0}_{rr}}.\label{eq47}
\end{equation}
Where, $g^{\prime}_{\mu\nu}$ is the metric with perturbation and $g^{\prime0}_{\mu\nu}$ is the metric without perturbation. We show numerical results in FIG. (5). We fix these parameters, Misner-Sharp quasi-local mass $m=1.0$, and the perturbation source star mass $M_{mp}=0.01m$, the perturbation source star orbit radius $R=30r_{H}$. One can find the tiny effects of perturbation on the trajectory of stars after they receive the perturbation, when the stars can move steadily relative to the wormhole without perturbation at first. In subsequent calculations we assume that the gravitational perturbation is continuous and invariant. A more general calculation can consider that the perturbation is time-dependent. From FIG. 5(a), we can find that the farther the star is from the wormhole, the less gravitational perturbation it receives. In FIG. 5(b), we have shown here that Perihelion and Aphelion in different eccentricity $\beta$ are affected by gravitational perturbation. But in fact, any position in trajectory would have a similar image. From FIG. 5(b), we can find that we need to use the difference in velocity $\Delta v$ of star to represent the change in redshift/blueshift, because the influence of gravitation perturbation is too small. 

\begin{figure*}[htbp]
	\centering
	\subfigure[The difference of radial-radial components' reciprocal as well as Eq. (\ref{eq47}) in region $r$ from $a$ to 10$a$ and the unit length of the abscissa axis in the figure is $a$. In the region near the wormhole, gravitational perturbation causes the radial-radial component to produce new negative branches. It leads to a larger difference in the region near the wormhole causing different physical scenarios.]{
		\includegraphics[width=0.45\linewidth]{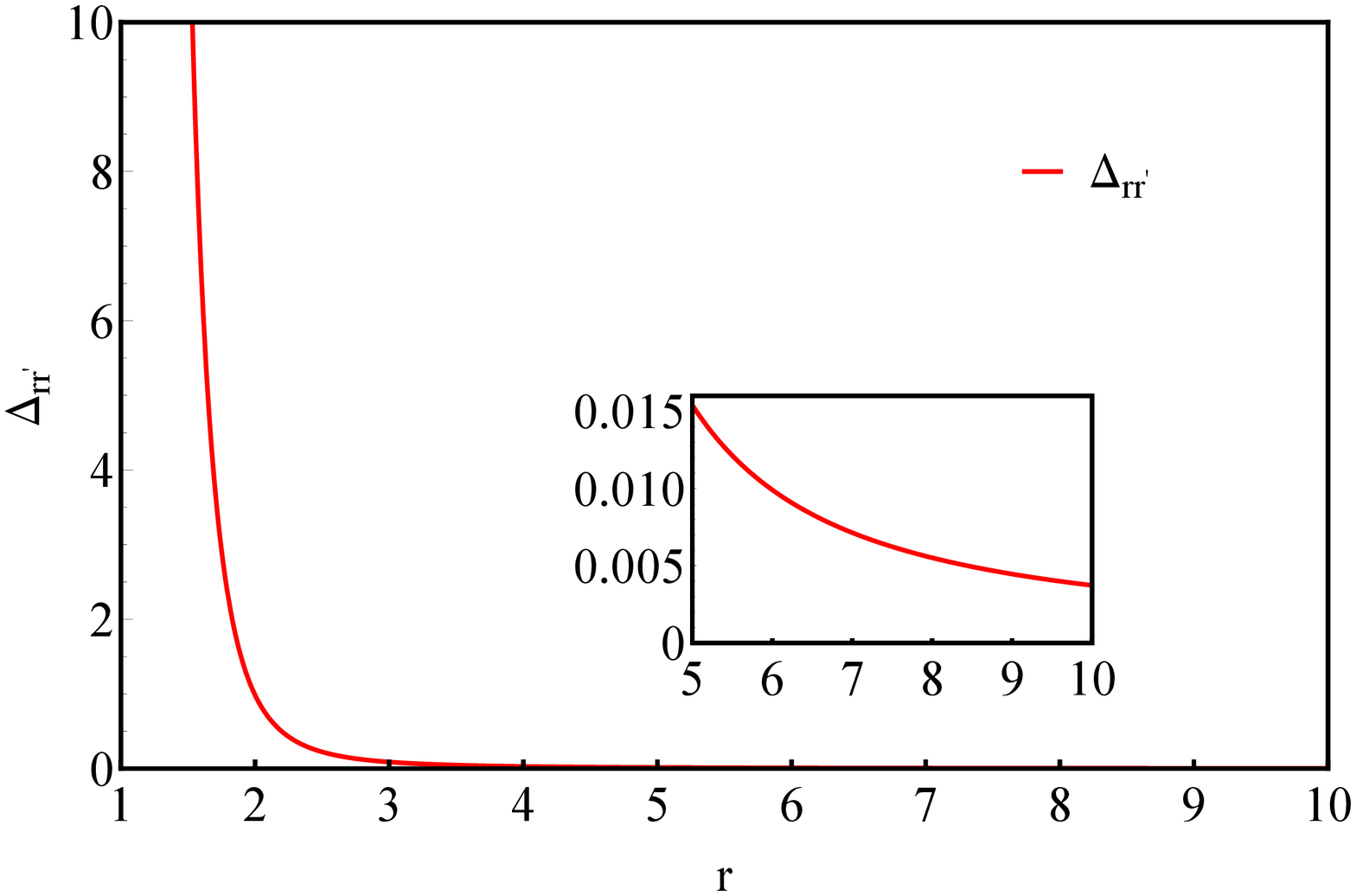}
	}
	\subfigure[The difference of radial-radial components' reciprocal as well as Eq. (\ref{eq47}) with different trajectory eccentricity $\beta$  from 0 to 1 at Perihelion and Aphelion. We use the orbital parameters in Section. II, and it can be seen that gravitational perturbation is very weak for our hypothetical trajectories.]{
		\includegraphics[width=0.48\linewidth]{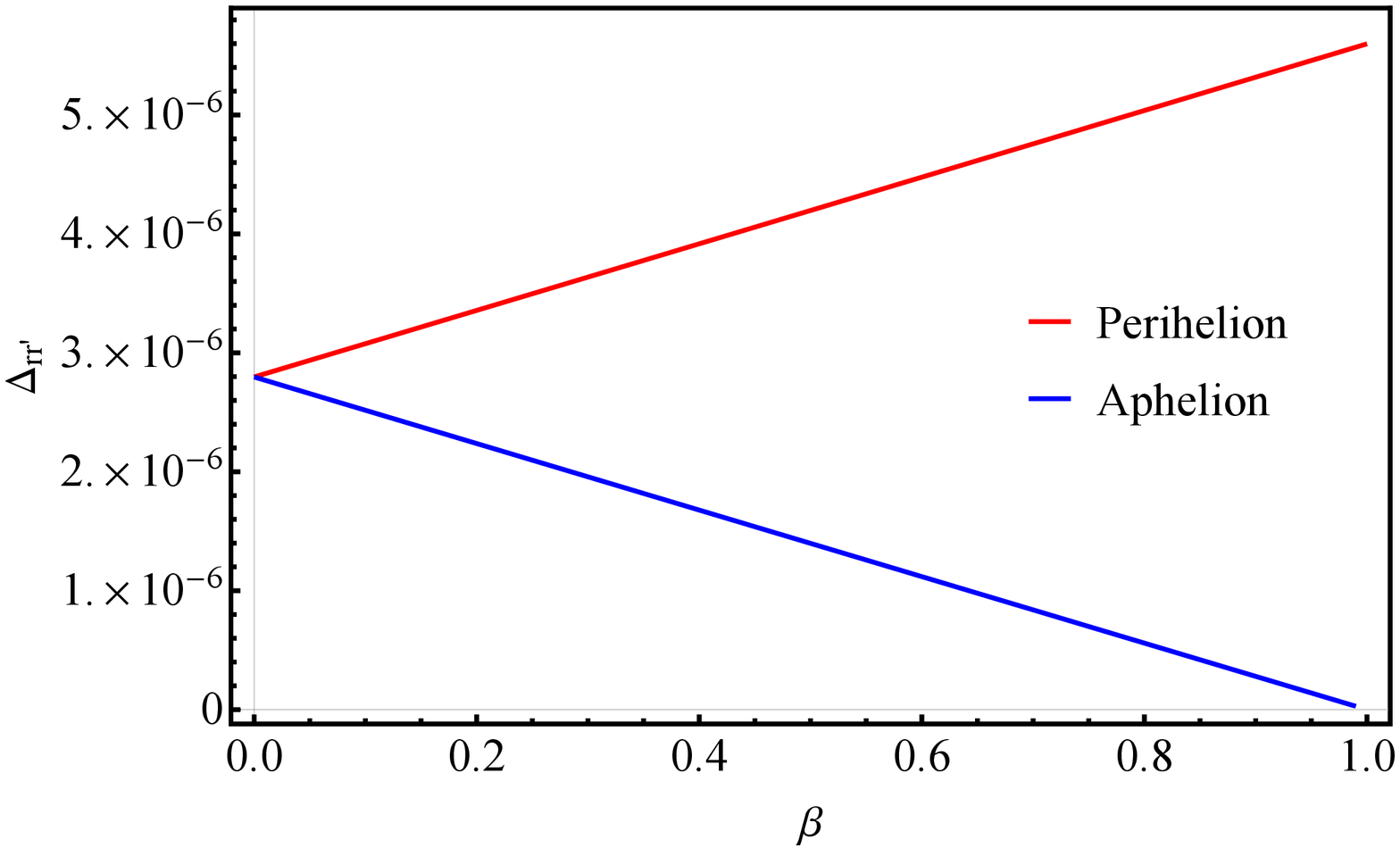}
	}
	\caption{The difference of metric variations as well as Eq. (\ref{eq47}) . In subfigure (a), we use the red solid line to show the difference in region $r\in[a,10a]$. In subfigure (b), we use the red solid line to signify the difference at Perihelion in region $\beta\in[0,1)$ and we use the red solid line to signify the difference at Aphelion in region $\beta\in[0,1)$.}
	\label{fig:5}
\end{figure*}

\begin{figure}[htbp]
	\centering
	\subfigure[The differences of constant trajectory displacement. As can be found from the figure that the differences are positive, indicating that gravitational perturbation suppresses the displacement.]{
		\includegraphics[width=1\linewidth]{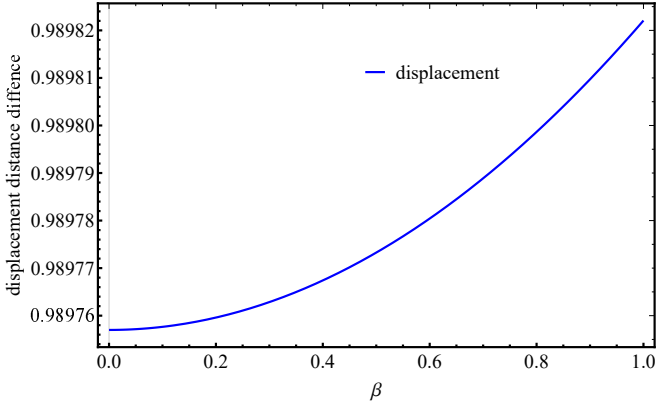}
	}
	\subfigure[The differences of oscillations around zero. From the figure, we can see that the amplitudes of the curves are still periodic but smaller. Hence, gravitational perturbation compresses the star's trajectory, but does not change its shape.]{
		\includegraphics[width=1\linewidth]{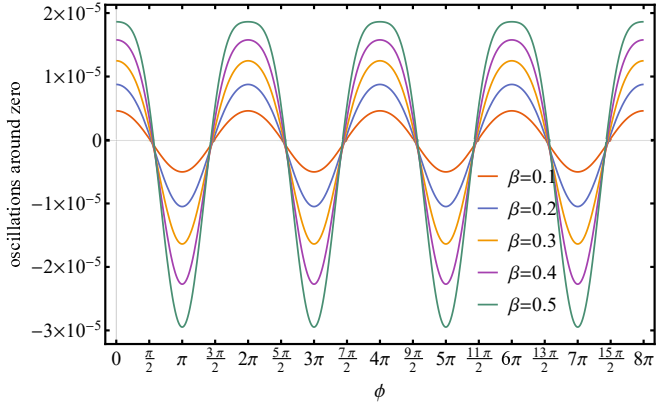}
	}
	\subfigure[The differences of accumulate over successive orbits of the long-axis revolves around the perihelion. From the figure, we can see that the amplitudes of the curves are cumulative but smaller too. Hence, gravitational perturbation also compresses the star's precession.]{
		\includegraphics[width=1\linewidth]{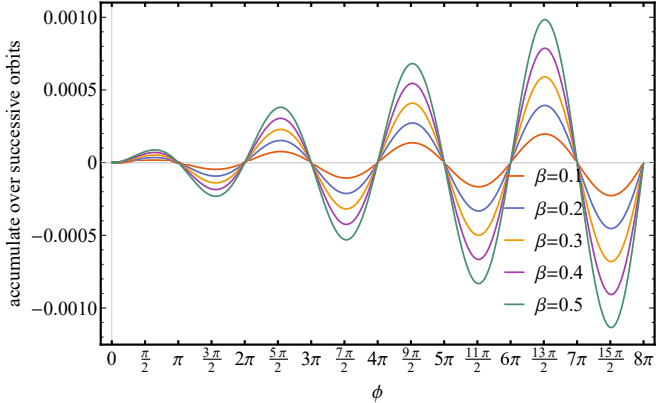}
	}
	\caption{The differences between $\tilde{A_1}, \tilde{B_1}$, $\tilde{C_1}$ in the Eq. (\ref{eq11})  and $\tilde{A_1}, \tilde{B_1}$, $\tilde{C_1}$ in the Eq. (\ref{eq49}) respectively. In subfigure (a), we use the blue solid line to show the change in displacement. In subfigre (b) and (c), we use five different colour solid lines to show the five different eccentricity $\beta=$ 0.1, 0.2, 0.3, 0.4 and 0.5 from top to bottom.}
	\label{fig:6}
\end{figure}

When we consider the actual orbits of celestial bodies, we can find $r\gg m$ in geometrized units. Then, one can obtain an approximate constant $\mathscr{H}$
\begin{equation}
	\begin{aligned}
		\mathscr{H}\approx\eta(\infty)&=\text{exp}\left[-\frac{2 \sqrt{m} }{\sqrt{2 a-m}}\arctan(\infty)\right]\\
		&=\text{exp}\left[-\frac{\pi\sqrt{m}}{\sqrt{2 a-m}}\right].
	\end{aligned}
\end{equation}
The approximate constant greatly reduces our calculations. Hence, we can rewrite the Eq. (\ref{eq45}) as
\begin{equation}
	\begin{aligned}
		\left(\frac{d r}{d \phi}\right)^{2}+\left(\frac{r^4}{L^2}+r^2\right)\left[1-\frac{2me^{-a/r}+2\mathscr{H}\kappa(R)}{r}\right]=\frac{E^2r^4}{L^2}.
	\end{aligned}
\end{equation}
Then the rest of the work is going to be similar to what we did in Sec. II, and finally we can find the equation of trajectory after perturbation
\begin{equation}
	\begin{aligned}
		x_0=&1+\beta^{\prime}\cos\phi,\\
		x_1=&\tilde{A_2}+\tilde{B_2}+\tilde{C_2}.\\
		\label{eq49}
	\end{aligned}
\end{equation}
Just like we discussed before, the solution of $x_1$ can be divided into three parts $\tilde{A_2}, \tilde{B_2}$ and $\tilde{C_2}$. The $\tilde{A_2}$ is simply a constant displacement, the $\tilde{B_2}$ is oscillations around zero and the $\tilde{C_2}$ is useful to accumulate over successive orbits. More details for the three parts, see Appendix I. Now, let us compare the changes of these three parts with perturbation and without perturbation. As discussed earlier, any position on the trajectory will be affected by gravitational perturbations, which have a cumulative effect on the kinematic shifts of photons. For simplifying the calculation, we choose Perihelion as the ``starting point'' for our calculations. Before the perturbation has propagated to reception-spacetime, the star moves periodically in its original trajectory. When the star moves at the Perihelion, we can obtain its angular velocity $d\phi/dt$ and tangential velocity $dr/dt$. At the same time, the perturbation propagates to reception-spacetime and keep stable. Then the star keep moving in the initial conditions $d\phi/dt$ and $dr/dt$. The energy and angular momentum of a star are conserved after it is disturbed by gravitational perturbations. From these assumptions we can obtain the angular momentum of star after receiving the perturbation
\begin{equation}
	\begin{aligned}
			\Omega&=\frac{d\phi}{dt}=\frac{d\phi}{d\tau}\frac{d\tau}{dt}\\
			&=\sqrt{-g_{tt}(r_a)}\frac{L}{r^2}=\sqrt{-g_{tt}^{\prime}(r_a)}\frac{L^{\prime}}{r^2},
	\end{aligned}
\end{equation}
so,
\begin{equation}
	L^{\prime}=\frac{\sqrt{-g_{tt}(r_a)}}{\sqrt{-g_{tt}^{\prime}(r_a)}}L,\label{eq50}
\end{equation}
where, $\Omega$ is the angular velocity, $r_a$ is the radius of Perihelion, $L$ and $g_{tt}$ are the angular momentum and metric component without perturbation, and then $L^{\prime}$ and $g_{tt}^{\prime}$ are the angular momentum and metric component with perturbation. In the same way, we can obtain the new eccentricity $\beta^{\prime}$ of the orbit 
\begin{equation}
		\frac{L^2}{m}\frac{1}{1+\beta}=r_a=\frac{L^{\prime2}}{m}\frac{1}{1+\beta^{\prime}},
\end{equation}
so,
\begin{equation}
	\begin{aligned}
		\beta^{\prime}&=\frac{L^{\prime2}(1+\beta)}{L^2}-1\\
		&=\frac{g_{tt}(r_a)}{g_{tt}^{\prime}(r_a)}\left(1+\beta\right)-1.\label{eq51}
	\end{aligned}
\end{equation}
We show the changes of these three parts as $\Delta\tilde{A}=\tilde{A_1}-\tilde{A_2}$, $\Delta\tilde{B}=\tilde{B_1}-\tilde{B_2}$ and $\Delta\tilde{C}=\tilde{C_1}-\tilde{C_2}$ separately in FIG. \ref{fig:6}. To plot FIG. \ref{fig:6}, we fix the Misner-Sharp quasi-local mass $m=1$, the star's angular momentum without perturbation $L=100a$, the eccentricity of the orbit without perturbation $\beta\in[0,1)$ and the rotation angle $\phi\in[0,8\pi]$. We can see from figure that the gravitational perturbation weakens the displacement behaviour of star, and at the same time suppresses the star's precession behaviour in the reception-spacetime.
Finally, we can get the polar equation of trajectory radius
\begin{equation}
	r=\frac{L^{\prime2}}{m}\frac{1}{1+\beta^{\prime}\cos\left(\phi-\epsilon^{\prime}\phi\right)}=\frac{(1-\beta^{\prime2})\alpha^{\prime}}{{1+\beta^{\prime}\cos\left(\phi-\epsilon^{\prime}\phi\right)}},
\end{equation}
where, $\alpha^{\prime}$ is the new semi-major axis, and $\epsilon^{\prime}$ is a parameter related to the precession angle
\begin{equation}
	\begin{aligned}
		\alpha^{\prime}&=\frac{1}{1-\beta^{\prime2}}\frac{L^{\prime2}}{m},\\
		\epsilon^{\prime}&=\frac{3 \mathscr{H} m \beta^{\prime}}{L^{\prime2}}+\frac{3 m^{2} \beta^{\prime}}{L^{\prime2}}-\frac{3 m^{2} \beta^{\prime}}{e L^{\prime2}}+\frac{6 m^{4} \beta^{\prime}}{e^{2} L^{\prime4}}-\frac{12 m^{4} \beta^{\prime}}{e L^{\prime4}}\\
		&\quad+\frac{8 m^{6} \beta^{\prime}}{e^{2} L^{\prime6}}+\frac{3 m^{4} \beta^{\prime3}}{2 e^{2} L^{\prime4}}-\frac{3 m^{4} \beta^{\prime3}}{e L^{\prime4}}+\frac{6 m^{6} \beta^{\prime3}}{e^{2} L^{\prime6}}\ll 1.
	\end{aligned}
\end{equation}

\begin{figure*}[htb]
	\centering
	\subfigure[The difference $\Delta z'$ between kinematic shifts of photons with perturbation and without perturbation. The photons are emitted by a star in particular trajectories with eccentricity $\beta$ from 0 to 0.5. The star moves away from our detector in the range of rotation angle $\phi$ from 0 to $\pi$. And the star moves toward to our detector in the range of rotation angle $\phi$ from $\pi$ to 2$\pi$.]{
		\includegraphics[width=0.5\linewidth]{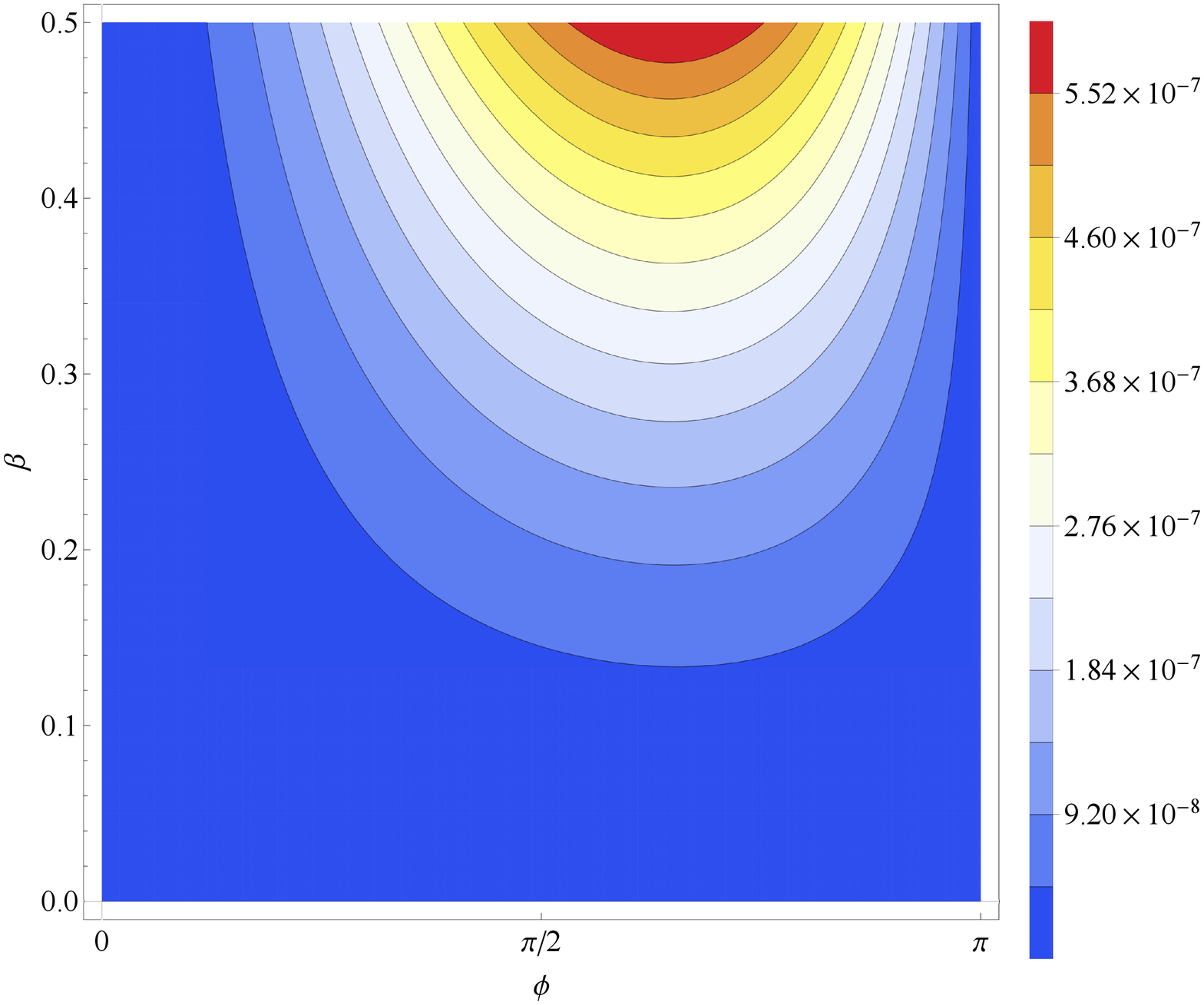} \includegraphics[width=0.5\linewidth]{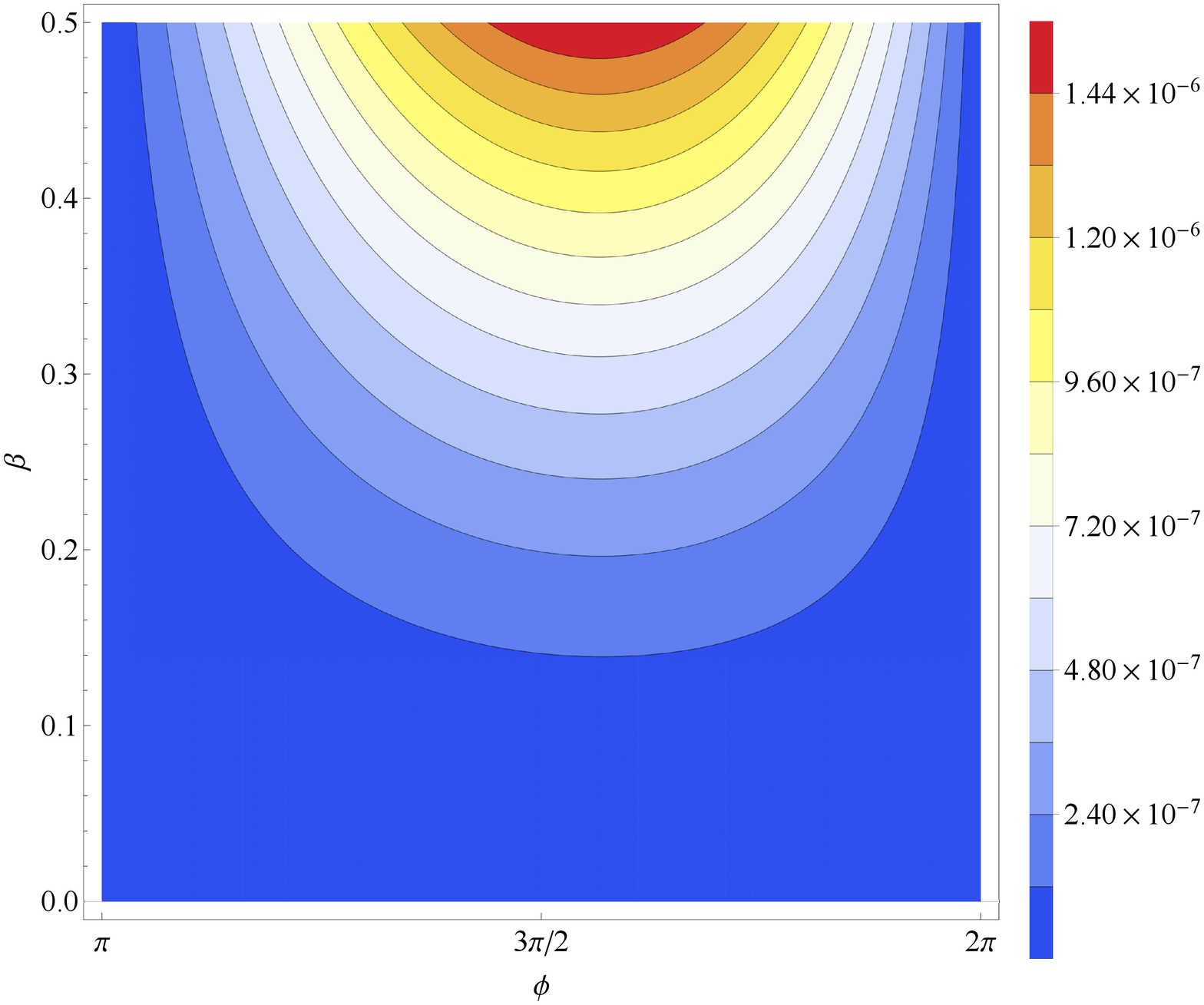}
	}
	\subfigure[The difference $\Delta z'$ between kinematic shifts of photons with perturbation and without perturbation. The photons are emitted by a star in particular trajectories with eccentricity $\beta$ from 0.5 to 1. The star moves away from our detector in the range of rotation angle $\phi$ from 0 to $\pi$. And the star moves toward to our detector in the range of rotation angle $\phi$ from $\pi$ to 2$\pi$.]{
		\includegraphics[width=0.5\linewidth]{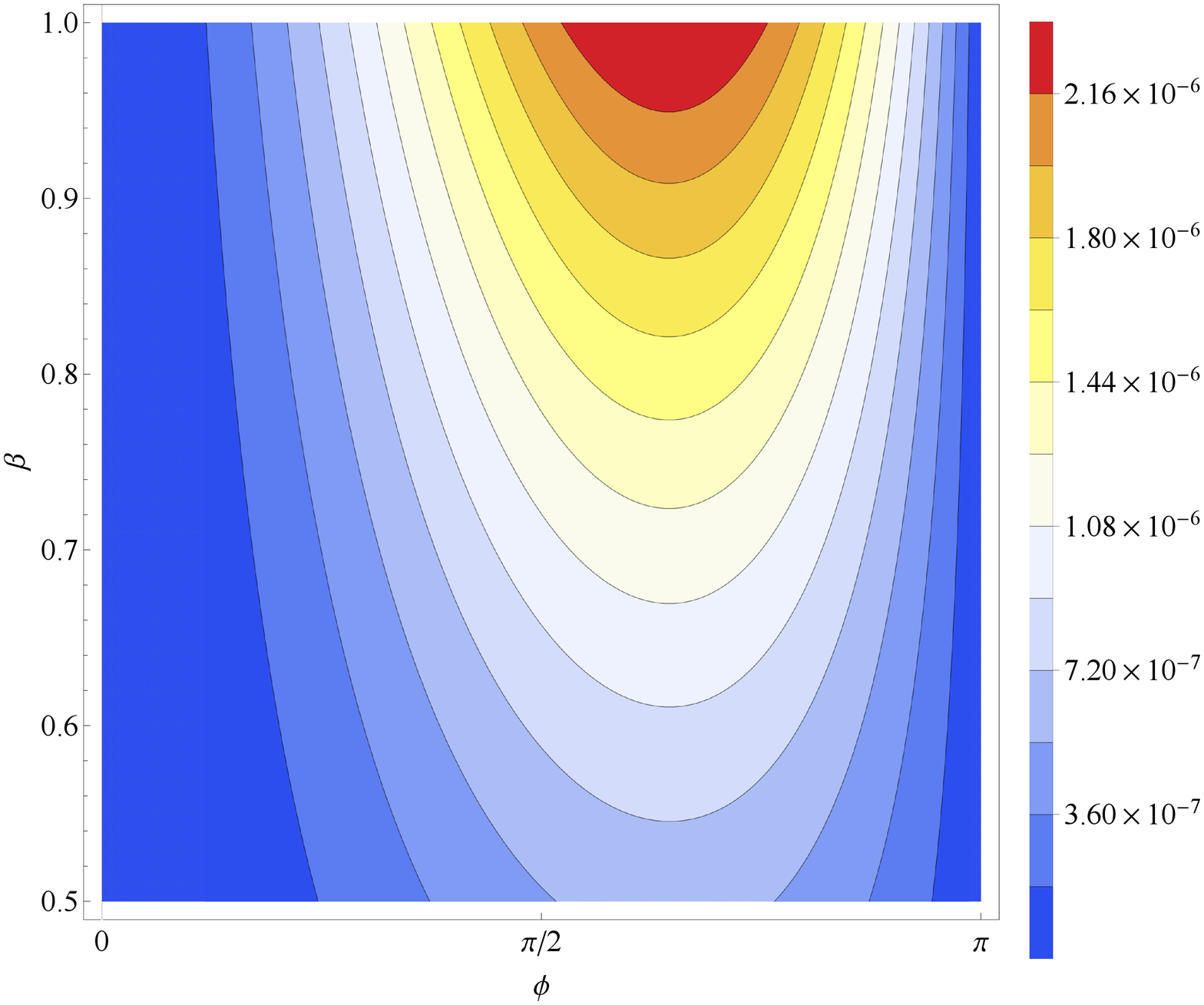} \includegraphics[width=0.5\linewidth]{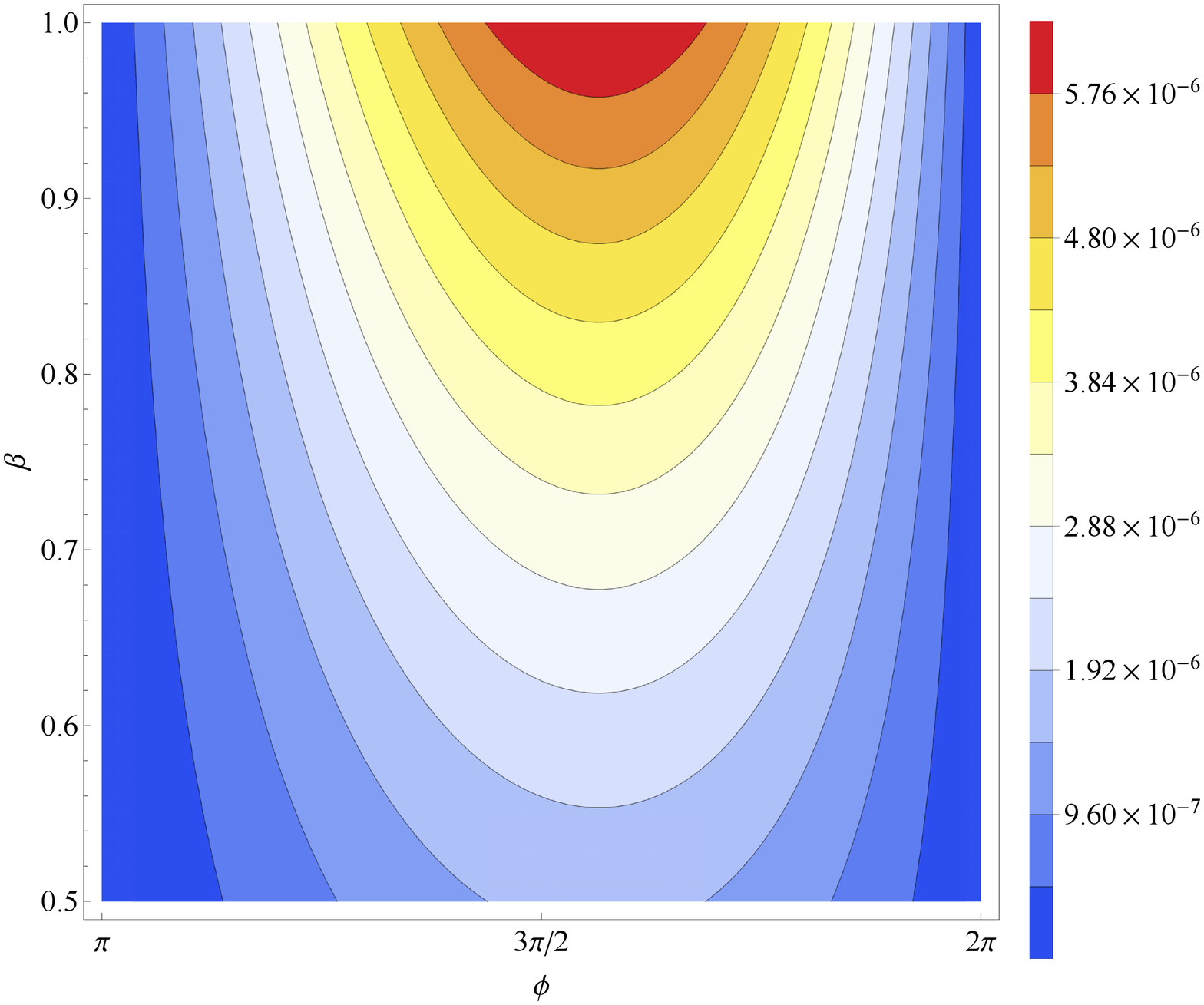}
	}
	\caption{The difference $\Delta z'$ between kinematic shifts of photons with perturbation and without perturbation in contour plot. The photons are emitted by a star in particular trajectories with eccentricity $\beta$ from 0 to 1 as well as the trajectory from circle, ellipse to parabola. The star moves away from our detector in the range of rotation angle $\phi$ from 0 to $\pi$. And the star moves toward to our detector in the range of rotation angle $\phi$ from $\pi$ to 2$\pi$.}
	\label{fig:7}
\end{figure*}

We use the same method as in Sec. II to describe the changes in frequency shifts of photons emitted by the star after reception-spacetime receives gravitational perturbations. However, we can see from FIG. \ref{fig:5} that the result of gravitational perturbation has become very tiny for the actual trajectory radius of the star. Therefore, we calculate the difference of redshift to describe the effect of gravitational perturbation
\begin{equation}
	\Delta z_{kin}=z_{kin}-z_{kin}^{\prime}.\label{eq54}
\end{equation}
Where, $z_{kin}$ is kinematic shift without perturbation shown as Eq. (\ref{eq20}), and $z_{kin}^{\prime}$ is kinematic shift with perturbation. To better present the results and facilitate discussion, we visualize Eq. (\ref{eq54}) as FIG. \ref{fig:7}. For drawing the FIG. \ref{fig:7}, we fix the Misner-Sharp quasi-local mass $m=1$. In reception-spacetime, the star's angular momentum without perturbation $L=100a$, the eccentricity of the orbit without perturbation $\beta\in[0,1)$ and the rotation angle $\phi\in[0,2\pi]$. The angular momentum will be changed after the gravitational perturbation transmitting to the space-time through the wormhole. Therefore, we assume that the star subjects to gravitational perturbation when it at the Perihelion, and the angular velocity of the star does not change. Then, the new angular momentum can be calculated by Eq. (\ref{eq50}). Similarly, The eccentricity of trajectory will also be changed after the gravitational perturbation transmits to the space-time through the wormhole. The new eccentricity $\beta^{\prime}$ can be obtained by Eq. (\ref{eq51}). In test-spacetime, we fix the perturbation source star mass $M_{mp}=0.01m$, it moves in equatorial plane along circular trajectory with the radius is $R=30r_H$. We use different colours to describe changes in the intensity of gravitational perturbation at photon frequency, as well as the differences of kinematic redshift/blueshift $\Delta z_{kin}$. The more colour tends to red, the more obvious effect of gravitational perturbation is. 

From FIG. \ref{fig:7}, we can find that with the increase of trajectory eccentricity, gravitational perturbation can modify the frequency redshift and blueshift of photons more obviously. And the effect of gravitational perturbation on the frequency blueshift of photons are more obvious than the effect of photon frequency redshift. FIG. \ref{fig:7} also tells us a very important information: under gravitational perturbation, the photon frequency kinematic shifts decrease, because $\Delta z_{kin}=z_{kin}-z_{kin}^{\prime}>0$. This is consistent with our discovery that gravitational perturbation inhibits or reduces the global translation, precession, angular momentum, and orbital eccentricity of the star. And these changes reflect the decrease of photon kinematic shifts

The slight differences in these shifts are the way for us to distinguish whether a star is moving around a black hole or wormhole. With the help of high-resolution and high-precision observation instruments on some larger scale sky surveys than before, we may be able to observe wormholes in the near future.

\section{Discussion and Final Remarks}

\label{sectionV}

In this paper, we show a method to distinguish between black holes and wormholes under the same space-time metric. The model starts from the nature of black holes and wormholes, and distinguishes wormholes from black holes by whether changes in the background gravity of test-spacetime will affect the motions of stars in reception-spacetime and changes in the frequencies of photons emitted by them. If the celestial body is a black hole, the gravitational perturbation in test-spacetime will not affect the motions of stars in reception-spacetime, and the frequencies of the photons received will not be changed. As the black hole does not connect the two copied space-time. But if the celestial body is a wormhole, the two space-times connected by the wormhole will interact with each other. The gravitational perturbation in test-spacetime will affect the motion of the star in reception-spacetime, and the frequencies of the photons received will also be changed.

We chose the regular space-time with asymptotically Minkowski core to construct the black hole and wormhole. We selected the $l=m=0$ perturbation mode in the Sec. III to calculate the gravitational perturbation result.

We found that under our chosen gravitational perturbation, the global translation and precession of the star in reception-spacetime will be suppressed, and the redshift and blueshift of the photon frequencies emitted by the star will also be inhibited. The small change in photon frequency is the key to distinguish a wormhole from black hole. With the improvement of the resolution and accuracy of the experiment, the small numerical difference will be shown within sight one day.

What is noteworthy is that the gravitational perturbation model used in this paper can also be extended to perturbation sources (massive stars) doing elliptical motion, scattering motion, or other more complex and closer to actual situations. Of course, the perturbation source can also be more than one, such as a group of stars, or any celestial body near the wormhole in the test-spacetime that produces the perturbation. In addition, we did not consider the influence of stars in the reception-spacetime on perturbation sources in the test-spacetime. This mutual influence may bring about a general result if one consider it.

\section*{Appendix I: The solutions of the radial equation of motion of the massive particle}
This part will show the details of the solutions of the radial equation of motion of the massive particle before and after perturbation. In the case which the reception-spacetime without perturbation, one can take the solutions of the zeroth-order part $x_0$ in the first-order part of Eq. (\ref{eq10}) to obtain the solutions $x_1$. The following are the three parts of the first-order of the solution $x_1$ which are divided by whether it contains the variable $\phi$ and whether the variable $\phi$ is multiplied by the trigonometric function. The segment which does not contain the variable $\phi$ is classified into $\tilde{A_1}$ regarded as a constant displacement. The segment which only has the trigonometric function of $\phi$ is classified into $\tilde{B_1}$ regarded as oscillating around zero. Furthermore, the segment which only has the variable $\phi$ multiplying the trigonometric function of $\phi$ is classified into $\tilde{C_1}$ regarded as accumulating over successive orbits. The following expressions are the three parts
\begin{equation}
	\begin{aligned}
		\tilde{A_1}=&1+\frac{3 m^{2}}{L^{2}}-\frac{3 m^{2}}{e L^{2}}+\frac{2 m^{4}}{e^{2} L^{4}}-\frac{8 m^{4}}{e L^{4}}+\frac{4 m^{6}}{e^{2}L^{6}}\\&+\frac{3m^{2} \beta^{2}}{2 L^{2}}+\frac{m^{4} \beta^{2}}{e^{2} L^{4}}-\frac{12m^{4} \beta^{2}}{e L^{4}}\\&+\frac{12 m^{6} \beta^{2}}{e^{2} L^{6}}+\frac{3 m^{6} \beta^{4}}{2 e^{2} L^{6}},
	\end{aligned}
\end{equation}
\begin{equation}
	\begin{aligned}
		\tilde{B_1}=&\frac{3 m^{2} \beta \cos\phi}{2 L^{2}}-\frac{3 m^{2} \beta \cos\phi}{4 e L^{2}}+\frac{m^{4} \beta \cos\phi}{e^{2}L^{4}}\\&-\frac{6 m^{4} \beta \cos\phi}{e L^{4}}+\frac{4 m^{6} \beta \cos\phi}{e^{2} L^{6}}-\frac{3 m^{4} \beta^{3} \cos\phi}{2 e L^{4}}\\&+\frac{3 m^{6} \beta^{3} \cos\phi}{e^{2}L^{6}}-\frac{m^{2} \beta^{2} \cos 2\phi}{2 L^{2}}-\frac{m^{4} \beta^{2} \cos 2\phi}{3 e^{2} L^{4}}\\&+\frac{4 m^{4} \beta^{2} \cos 2\phi}{e L^{4}}-\frac{4 m^{6} \beta^{2} \cos 2\phi}{e^{2} L^{6}}\\&-\frac{2 m^{6} \beta^{4} \cos 2\phi}{3 e^{2} L^{6}}+\frac{m^{4} \beta^{3} \cos 3\phi}{4 eL^{4}}\\&-\frac{m^{6} \beta^{3} \cos3\phi}{2 e^{2} L^{6}}-\frac{m^{6} \beta^{4} \cos 4\phi}{30 e^{2} L^{6}},
	\end{aligned}
\end{equation}
\begin{equation}
	\begin{aligned}
		\tilde{C_1}=&\frac{3 m^{2} \beta \phi \sin \phi}{L^{2}}-\frac{3 m^{2} \beta \phi \sin \phi}{2 e L^{2}}+\frac{2 m^{4} \beta \phi \sin \phi}{e^{2} L^{4}}\\&-\frac{12 m^{4} \beta \phi \sin \phi}{e L^{4}}+\frac{8 m^{6} \beta \phi \sin \phi}{e^{2} L^{6}}\\&-\frac{3 m^{4} \beta^{3} \phi \sin \phi}{e L^{4}}+\frac{6 m^{6} \beta^{3} \phi \sin \phi}{e^{2} L^{6}}.
	\end{aligned}
\end{equation}

In the case which the reception-spacetime with perturbation, one can take the solutions of the zeroth-order part $x_0$ in the first-order part of Eq. (\ref{eq49}) to obtain the solutions $x_1$. The following are the three parts of the first-order of the solution $x_1$ which are divided by whether it contains the variable $\phi$ and whether the variable $\phi$ is multiplied by the trigonometric function. The segment which does not contain the variable $\phi$ is classified into $\tilde{A_2}$ regarded as a constant displacement. The segment which only has the trigonometric function of $\phi$ is classified into $\tilde{B_2}$ regarded as oscillating around zero. Furthermore, the segment which only has the variable $\phi$ multiplying the trigonometric function of $\phi$ is classified into $\tilde{C_2}$ regarded as accumulating over successive orbits. The following expressions are the three parts
\begin{equation}
	\begin{aligned}
		\tilde{A_2}=&\frac{\mathscr{H}}{m}+\frac{3 \mathscr{H} m}{L^{\prime2}}+\frac{3 m^{2}}{L^{\prime2}}-\frac{4 m^{2}}{e L^{\prime2}}+\frac{4 m^{4}}{e^{2} L^{\prime4}}\\&-\frac{8 m^{4}}{e L^{\prime4}}+\frac{4 m^{6}}{e^{2} L^{\prime6}}+\frac{3 \mathscr{H} m \beta^{\prime2}}{2 L^{\prime2}}+\frac{3 m^{2} \beta^{\prime2}}{2 L^{\prime2}}\\&-\frac{m^{2} \beta^{\prime2}}{e L^{\prime2}}+\frac{6 m^{4} \beta^{\prime2}}{e^{2} L^{\prime4}}-\frac{12 m^{4} \beta^{\prime2}}{e L^{\prime4}}\\&+\frac{12 m^{6} \beta^{\prime2}}{e^{2} L^{\prime6}}+\frac{3 m^{6} \beta^{\prime4}}{2 e^{2} L^{\prime6}},
	\end{aligned}
\end{equation}
\begin{equation}
	\begin{aligned}
		\tilde{B_2}=&\frac{3 \mathscr{H}m \beta^{\prime} \cos \phi}{2 L^{\prime2}}+\frac{3 m^{2} \beta^{\prime} \cos \phi}{2 L^{\prime2}}-\frac{3 m^{2} \beta^{\prime}\cos \phi}{2 e L^{\prime2}}\\&+\frac{3 m^{4} \beta^{\prime} \cos\phi}{e^{2} L^{\prime4}}-\frac{6 m^{4}\beta^{\prime} \cos\phi}{e L^{\prime4}}+\frac{4 m^{6}\beta^{\prime}\cos\phi}{e^{2} L^{\prime6}}\\&+\frac{3 m^{4} \beta^{\prime3} \cos \phi}{4 e^{2} L^{\prime4}}-\frac{3 m^{4} \beta^{\prime3} \cos \phi}{2 e L^{\prime4}}\\&+\frac{3 m^{6} \beta^{\prime3} \cos \phi}{e^{2} L^{\prime6}}-\frac{\mathscr{H} m \beta^{\prime2} \cos 2\phi}{2 L^{\prime2}}\\&-\frac{m^{2} \beta^{\prime2} \cos 2\phi}{2 L^{\prime2}}+\frac{m^{2} \beta^{\prime2} \cos 2\phi}{3 e L^{\prime2}}\\
		&-\frac{2 m^{4} \beta^{\prime2} \cos 2\phi}{e^{2} L^{\prime4}}+\frac{4 m^{4} \beta^{\prime2} \cos 2\phi}{e L^{\prime4}}\\&-\frac{4 m^{6} \beta^{2} \cos 2\phi}{e^{2} L^{\prime6}}-\frac{2 m^{6} \beta^{\prime4} \cos 2\phi}{3 e^{2} L^{\prime6}}\\&-\frac{m^{4} \beta^{\prime3} \cos 3\phi}{8 e^{2} L^{\prime4}}+\frac{m^{4} \beta^{\prime3} \cos 3\phi}{4 eL^{\prime4}}\\
		&-\frac{m^{6} \beta^{\prime3} \cos 3\phi}{2 e^{2} L^{\prime6}}-\frac{m^{6} \beta^{\prime4} \cos 4\phi}{30 e^{2} L^{\prime6}},
	\end{aligned}
\end{equation}
\begin{equation}
	\begin{aligned}
		\tilde{C_2}=&\frac{3 \mathscr{H} m \beta^{\prime} \phi \sin \phi}{L^{\prime2}}+\frac{3 m^{2} \beta^{\prime} \phi \sin \phi}{L^{\prime2}}\\&-\frac{3 m^{2}\beta^{\prime} \phi \sin \phi}{e L^{\prime2}}+\frac{6 m^{4}\beta^{\prime}\phi \sin \phi}{e^{2} L^{\prime4}}\\&-\frac{12 m^{4} \beta^{\prime} \phi \sin \phi}{e L^{\prime4}}+\frac{8 m^{6}\beta^{\prime}\phi \sin \phi}{e^{2} L^{\prime6}}\\&+\frac{3 m^{4} \beta^{\prime3} \phi \sin \phi}{2 e^{2} L^{\prime4}}-\frac{3 m^{4} \beta^{\prime3} \phi \sin \phi}{e L^{\prime4}}\\&+\frac{6 m^{6} \beta^{\prime3} \phi \sin \phi}{e^{2} L^{\prime6}}.
	\end{aligned}
\end{equation}

\section*{Appendix II: Gauge invariant perturbation details}
\subsection{The brief introduction of the gauge invariant perturbation}

Generally speaking, it is very complicated to solve the exact solution of the Einstein field equation. Therefore, starting with the exact solution $\left(M, g\right)$ of the available field equation and modifying it by using the perturbation theory to obtain a new solution $\left(M', g'\right)$  is also a way to solve the field equation, where the $M, M'$ refer to manifold and the $g, g'$ refer to metric. The solution of a field equation that can be called an exact solution is a complete atlas of coordinate charts that can be used to describe the changes of various quantities on the manifold $M$. These coordinate charts currently only map from a subset of $\mathbb{R}^{4}$ to $M$, and we hope to apply these same coordinates to the physical manifold $M'$ by some operations. In fact, we can accomplish this work through a map $\phi: M\rightarrow M'$. Since we may wish to extend a number of smoothly related coordinates and every point in $M'$ should have its own coordinate labelling which means that no two points have the same coordinates, the map $\phi$ would be a smooth bijection called diffeomorphism. The diffeomorphism allows us to compare tensors of different points with the same coordinate values. Given the coordinate system on the background manifold $M$, the diffeomorphism smoothly assigns the same coordinate value between the points of the physical manifold $M'$. Under our selected physical spacetime, $(M', g')$ is only slightly different from background spacetime $\left(M, g\right)$, and $\phi$ tells us which points will be treated as the same point, such as $f_{\phi}:=\phi^{*} g'-g$, where the $\phi^{*}$ is the pullback. The value of any tensor or tensor perturbation usually depends on the specific correspondence between $M$ and $M'$, but there is no general preferred mapping $\phi$ between any two manifolds. This leads us to arbitrarily choose the mapping without changing the physical situation. Therefore, the selection of mapping $\phi$ is called gauge choice. We can choose any gauge as long as all equations are formed in terms of physical quantities which do not depend on the choice of gauge. They are known as the gauge freedom of perturbation theory. The gauge independent physical quantities are known as gauge invariants.

As mentioned above, the $\left(M, g\right)$ and $(M', g')$ are slightly different between each other then the $f_{\phi}$ is small everywhere under the gauge $\phi$. However, one can not ensure that $f_{\phi}$ will remain small in a different gauge $\varphi$, since $\varphi^{-1}$ could map to a point where $g$ is different. Luckily, the different gauges referring to different points of the background manifold can be written as
\begin{equation}
	\begin{aligned}
		\psi&: M_{0} \rightarrow M_{0}, \\
		\psi&=\varphi^{-1} \circ \phi.
	\end{aligned}
\end{equation}
Note that $\psi$ is a diffeomorphism from the background manifold to itself, and that $\phi \circ \psi^{-1}=\varphi$. The diffeomorphism $\psi$ can then be used to change from one gauge to another. Hence, one can obtain the perturbation of the metric
\begin{equation}
	f_{\varphi}=\varphi^{*} g-g'=\psi_{*} \phi^{*} g-g'.
\end{equation}
Where, the $\psi_{*}$ is the pushforward. The perturbation $f_{\phi}$ and $f_{\varphi}$ must can be compared at the same point, so one can apply the pushforward $\psi_{*}$ to $f_{\phi}$
\begin{equation}
	\Delta f_{\psi} \equiv f_{\varphi}-\psi_{*} f_{\phi}=\psi_{*} g-g,\label{eq73}
\end{equation}
which depends only on $\psi$ and the background metric. This equation describes how the perturbation on the background changes with a gauge transformation. But there is a question: when the change is very small, whether the perturbation remains very small. Therefore, we need to talk about the infinitesimal diffeomorphisms. The infinitesimal diffeomorphisms are generated by vector fields, so we suppose $\psi$ is an infinitesimal diffeomorphism generated by a vector field $\xi^{a}$. The change of any tensor field under an infinitesimal pushforward $\psi_{*}$ is the Lie derivative of that tensor field with respect to $\xi^{a}$. Therefore, the change of the perturbation of Eq. (\ref{eq73}) is
\begin{equation}
	\Delta f_{\psi}=\mathscr{L}_{\xi} g=\nabla_{a} \xi_{b}+\nabla_{b} \xi_{a},\label{eq74}
\end{equation}
where $\xi$ is the gauge vector. One can find that the right of Eq. (\ref{eq74}) is not zero for an arbitrary vector field, so the components of $f_{\psi}$ are variable under a choice of gauge. Since we need that any physical quantity we used is independent of a choice of gauge, it is necessary to find some gauge invariants.

One possible natural definition of a gauge invariant quantity is the tensor field $T$ on $M'$, and then the components of $T$ transformed to the coordinates induced by $\phi$ are the same for all gauges. However, this definition is too restrictive on the quantities $T$ which must be either vanishing constant scalar fields or tensors and the $\phi^{*} T$ is a tensor field on the background $M$ which we do not always care about since we consider the physical quantities in the physical manifold $M'$ \cite{Sachs:1964,Bruni:1999et,Chen:2016plo}. Moreover, the most important is that we need the quantities that are formed out of selective perturbations of tensor fields, rather than the tensor fields themselves. Stewart and Walker researched the gauge invariance of the perturbations firstly \cite{Stewart:1974uz}. They found that the perturbation $\Delta Q$ of one quantity $Q$ is gauge invariant if the value of the quantity itself equals to zero on the background manifold $M$. Afterwards, Bardeen \cite{Bardeen:1980kt} adopted a method to calculate the gauge invariants whose core idea is starting in an arbitrary gauge $\phi$ and showing the existence of the corresponding gauge vector $\xi$ which transforms the tensor perturbation to one of these useful gauges from the Eq. (\ref{eq74}). In the Sec. III, we show the calculating process under setting the desired components of the metric to their restricted values.

\subsection{The normalization functions and the $A-K$ decomposition}
The normalizations functions in Eq. (\ref{eq30}) can be obtained by projecting each pure-spin vector and tensor harmonic into itself over the 2-sphere. For example, the pure-spin vector $Y_{a}^{R, \ell' m'}$
\begin{equation}
	\begin{aligned}
		&\oint Y_{a}^{R, \ell' m'}\left(Y_{R, \ell'' m''}^{a}\right)^{*} \mathrm{d} \Omega\\
		=&\oint n_{a}Y^{\ell' m'}(n^{a}Y_{\ell'' m''})^{*}\sin\theta d\theta d\phi\\
		=&\left(1-\frac{2 m e^{-a / r}}{r}\right)\delta_{\ell' \ell''} \delta_{m' m''},
	\end{aligned}
\end{equation}
and for the pure-spin tensor $T_{a b}^{T 0, \ell' m'}$
\begin{equation}
	\begin{aligned} 
		&\oint T_{a b}^{T 0, \ell' m'}\left(T_{T0, \ell'' m''}^{a b}\right)^{*} \mathrm{d} \Omega\\
		=&\oint \sigma_{a b}Y^{\ell' m'}(\sigma^{ab}Y_{\ell'' m''})^{*}\sin\theta d\theta d\phi\\
		=&\oint\left(r^2*\frac{1}{r^2}+r^2\sin^2\theta*\frac{1}{r^2\sin^2\theta}\right)Y^{\ell' m'}Y^{\ell'' m''}\sin\theta d\theta d\phi\\
		=&2\delta_{\ell' \ell''} \delta_{m' m''}.
	\end{aligned}
\end{equation}
Then, the full expression of normalizations functions in Eq. (30) are listed below
\begin{equation}
	\begin{aligned}
		&\oint Y_{a}^{B, \ell' m'}\left(Y_{B, \ell'' m''}^{a}\right)^{*} \mathrm{d} \Omega=\ell'(\ell'+1) \delta_{\ell' \ell''} \delta_{m' m''},\\
		&\oint Y_{a}^{E, \ell' m'}\left(Y_{E, \ell'' m''}^{a}\right)^{*} \mathrm{d} \Omega=\ell'(\ell'+1) \delta_{\ell' \ell''} \delta_{m' m''},\\
		&\oint Y_{a}^{R, \ell' m'}\left(Y_{R, \ell'' m''}^{a}\right)^{*} \mathrm{d} \Omega=\left(1-\frac{2 m e^{-a / r}}{r}\right) \delta_{\ell' \ell''} \delta_{m' m''},\\
		&\oint T_{a b}^{L 0, \ell' m'}\left(T_{L 0, \ell'' m''}^{a b}\right)^{*} \mathrm{d} \Omega=\left(1-\frac{2 m e^{-a / r}}{r}\right)^{2} \delta_{\ell' \ell''} \delta_{m' m''},\\
		&\oint T_{a b}^{T 0, \ell' m'}\left(T_{T 0, \ell'' m''}^{a b}\right)^{*} \mathrm{d} \Omega=2 \delta_{\ell' \ell''} \delta_{m' m''},\\
		&\oint T_{a b}^{B 1, \ell' m'}\left(T_{B 1, \ell'' m''}^{a b}\right)^{*} \mathrm{d} \Omega\\&=\left(1-\frac{2 m e^{-a / r}}{r}\right) \frac{\ell'(\ell'+1)}{2} \delta_{\ell' \ell''} \delta_{m' m''},\\
		&\oint T_{a b}^{B 2, \ell' m'}\left(T_{B 2, \ell'' m''}^{a b}\right)^{*} \mathrm{d} \Omega=\frac{(\ell'+2) !}{2(\ell'-2) !} \delta_{\ell' \ell''} \delta_{m' m''},\\
		&\oint T_{a b}^{E 1, \ell' m'}\left(T_{E 1, \ell'' m''}^{a b}\right)^{*} \mathrm{d} \Omega\\&=\left(1-\frac{2 m e^{-a / r}}{r}\right) \frac{\ell'(\ell'+1)}{2} \delta_{\ell' \ell''} \delta_{m' m''},\\
		&\oint T_{a b}^{E 2, \ell' m'}\left(T_{E 2, \ell'' m''}^{a b}\right)^{*} \mathrm{d} \Omega=\frac{(\ell'+2) !}{2(\ell'-2) !} \delta_{\ell' \ell''} \delta_{m' m''}.\\
	\end{aligned}
\end{equation}

Similarly, the $A-K$ components of Eq. (\ref{eq27}) can be obtained by projecting themselves onto each associated vector or tensor harmonic then the expression can recover to $h_{ab}$. Take the component $A$ for example, its associate vector harmonic from Eq. (\ref{eq27}) is $v_{a}v_{b}Y^{\ell' m'}$, hence
\begin{equation}
		\oint Av_{a}v_{b}Y^{\ell' m'} \mathrm{d} \Omega=h_{ab}.
\end{equation}
Then,
\begin{equation}
A=\tilde{N(r,\ell')}\oint v^{a}v^{b}h_{ab}Y_{\ell'' m''}^{*} \mathrm{d} \Omega,
\end{equation}
where $\tilde{N(r,\ell')}$ is the specific normalization factor of the harmonic $v_{a}v_{b}Y^{\ell' m'}$, it can be found through
\begin{equation}
\frac{1}{\tilde{N(r,\ell')}}\oint v_{a}v_{b}v^{a}v^{b}Y^{\ell' m'}Y_{\ell'' m''}^{*}\mathrm{d} \Omega=\delta_{\ell' \ell''} \delta_{m' m''},\\
\end{equation}
then,
\begin{equation}
\tilde{N(r,\ell')}=\left(1-\frac{2 m e^{-a / r}}{r}\right)^{2}.
\end{equation}
Therefore, the component $A$ can be written as
\begin{equation}
A=\left(1-\frac{2 m e^{-a / r}}{r}\right)^{2}\oint v^{a}v^{b}h_{ab}Y_{\ell' m'}^{*} \mathrm{d} \Omega.
\end{equation}
Where, for writing convenience, we will write $\ell'', m''$ as $\ell', m'$ without causing ambiguity. Employing the same method, we can obtain the whole $A-K$ components of the perturbation metric $h_{ab}$
\begin{equation}
	\begin{aligned}
		A&=\left(1-\frac{2 m e^{-a / r}}{r}\right)^{2} \oint v^{a} v^{b} h_{a b} Y_{\ell' m'}^{*} \mathrm{d} \Omega,\\
		B&=\frac{-\left(1-\frac{2 m e^{-a / r}}{r}\right)}{\ell'(\ell'+1)} \oint v^{a} Y_{E}^{b *} h_{a b} \mathrm{d} \Omega,\\
		C&=\frac{-\left(1-\frac{2 m e^{-a / r}}{r}\right)}{\ell'(\ell'+1)} \oint v^{a} Y_{B}^{b *} h_{a b} \mathrm{d} \Omega,\\
		D&=-\oint v^{a} Y_{R}^{b *} h_{a b} \mathrm{d} \Omega,\\
		E&=\frac{1}{2} \oint T_{T 0}^{a b *} h_{a b} \mathrm{d} \Omega, \quad F=\frac{2(l-2) !}{(\ell'+2) !} \oint T_{E 2}^{a b *} h_{a b} \mathrm{d} \Omega,\\ 
		G&=\frac{2(\ell'-2) !}{(\ell'+2) !} \oint T_{B 2}^{a b *} h_{a b} \mathrm{d} \Omega,\\ 
		H&=\frac{(\ell'-1) !}{(\ell'+1) !}\left(1-\frac{2 m e^{-a / r}}{r}\right)^{-1} \oint T_{E 1}^{a b *} h_{a b} \mathrm{d} \Omega,\\
		J&=\frac{(\ell'-1) !}{(\ell'+1) !}\left(1-\frac{2 m e^{-a / r}}{r}\right)^{-1} \oint T_{B 1}^{a b *} h_{a b} \mathrm{d} \Omega,\\
		K&=\left(1-\frac{2 m e^{-a / r}}{r}\right)^{-2} \oint T_{L 0}^{a b *} h_{a b} \mathrm{d} \Omega.
	\end{aligned}
\end{equation}

\subsection{The general approach for gauge invariants}
This part is going to show how we can get gauge invariants from a gauge transformation. Consider a infinitesimal diffeomorphisms generated by a vector field $\xi^{a}$, the first-order metric perturbation $h_{ab}$ is changed under the form of the Eq. (\ref{eq74})
\begin{equation}
	h_{a b}^{\text {new}}=h_{a b}^{\text {old}}-2 \nabla_{(a} \xi_{b)},
\end{equation}
where, $2 \nabla_{(a} \xi_{b)}=\nabla_{a} \xi_{b}+\nabla_{b} \xi_{a}$. Since we consider our perturbation source moves in a circular trajectory, the parameters $\ell', \ell'', m', m''$ are all set to zero, then we can obtain a simply form of $h_{ab}$ which is shown as Eq. (\ref{eq32}). After this, we can decompose the gauge vector into the pure-spin harmonic basis
\begin{equation}
	\xi_{a}=\mathrm{P} v_{a} Y_{\ell' m'}+\mathrm{R} n_{a} Y_{\ell' m'}=\frac{1}{2 \sqrt{\pi}}\left(\mathrm{P} v_{a}+\mathrm{R} n_{a}\right).
\end{equation}
The symbols $\mathrm{P}$ and $\mathrm{R}$ represent two scalar functions of $(t, r)$ with harmonic labels and coordinate dependence is suppressed for convenience. The functions $\mathrm{P}$ and $\mathrm{R}$ describe the two degrees of gauge freedom. Then we can calculate the $A-K$ term of $2 \nabla_{(a} \xi_{b)}$. Take the component $A$ for example
\begin{equation}
	\begin{aligned}
		\Delta \mathrm{A}&\equiv 2\left(1-\frac{2 m e^{-a / r}}{r}\right)^{2} \oint v^{a} v^{b} \nabla_{a} \xi_{b} Y_{0 0}^{*} \mathrm{d} \Omega\\
		&=\frac{2}{4\pi}\left(1-\frac{2 m e^{-a / r}}{r}\right)^{2} \oint v^{a} v^{b}\nabla_{a}\left(\mathrm{P} v_{b}+\mathrm{R} n_{b}\right) \mathrm{d} \Omega\\
		&=-2\frac{\partial}{\partial t}\mathrm{P}-\frac{\left[2 m e^{-\frac{2 a}{r}} (r-a) \left(r e^{a/r}-2 m\right)\right]}{r^4}\mathrm{R}.
	\end{aligned}
\end{equation}
This term $\Delta \mathrm{A}$ alone is responsible for changes to the component $A$ of perturbation metric $h_{a b}$
\begin{equation}
	\mathrm{A}_{\text {new}}=\mathrm{A}_{\text{old}}-\Delta \mathrm{A}.
\end{equation}
The ``new" and ``old" subscripts correspond to projections of $h_{a b}^{\text {new}}$ and $h_{a b}^{\text {old}}$, respectively. Moreover, using the same method, we can find the components $D$, $E$, $K$. We list them in the below
\begin{equation}
	\begin{aligned}
		&\Delta \mathrm{A}=-2\frac{\partial}{\partial t}\mathrm{P}-\frac{\left[2 m e^{-\frac{2 a}{r}} (r-a) \left(r e^{a/r}-2 m\right)\right]}{r^4}\mathrm{R},\\
		&\Delta \mathrm{D}=-\frac{2 m (r-a)}{r^2 \left(r e^{a/r}-2 m\right)}\mathrm{P}-2\frac{\partial}{\partial t}\mathrm{R},\\
		&\Delta \mathrm{E}=\frac{2 \left(r-2 m e^{-a/r}\right)}{r^2}\mathrm{R},\\
		&\Delta \mathrm{K}=2\left[\frac{\partial }{\partial r}+\frac{m (r-a)}{r^2 \left(r e^{a/r}-2 m\right)}\right]\mathrm{R}.\label{eq86}
	\end{aligned}
\end{equation}

As we have the gauge transformation on the metric projections, we can start to find a class of gauge invariant quantities in the used spacetime. The approach used to find gauge invariants in the spacetime below follows in a similar manner to that of Gerlach and Sengupta \cite{Gerlach:1979rw}. Further discussion of the Gerlach and Sengupta decomposition and gauge choices may be found in Brizuela et al \cite{Brizuela:2007zza}. One may find that Eq. (\ref{eq86}) can be inverted to find the components of $\xi^{a}$ and their derivatives in terms of changes in the metric under the gauge transformation. Let us take the most obvious components $\Delta \mathrm{E}$ and $\Delta \mathrm{K}$ for example, from
\begin{equation}
		\Delta \mathrm{E}=\frac{2\left(r-2 m e^{-a/r}\right)}{r^2}\mathrm{R}\\
\end{equation}
one can obtain
\begin{equation}
\mathrm{R}= \frac{r^2}{2 \left(r-2 m e^{-a/r}\right)}\Delta \mathrm{E}.
\end{equation}
We can take the expression of $\mathrm{R}$ into $\Delta \mathrm{K}$
\begin{equation}
	\begin{aligned}
		\Delta \mathrm{K}&=2\left[\frac{\partial }{\partial r}+\frac{m (r-a)}{r^2 \left(r e^{a/r}-2 m\right)}\right]\frac{r^2}{2 \left(r-2 m e^{-a/r}\right)}\Delta \mathrm{E}\\
		&=\frac{e^{a/r}\left[r\left(e^{a/r}-3 m\right)+am\right]}{\left(r e^{a/r}-2 m\right)^2}\Delta \mathrm{E}+\frac{r^2 e^{a/r}}{\left(r e^{a/r}-2 m\right)}\frac{\partial }{\partial r}\Delta \mathrm{E}.\\
	\end{aligned}
\end{equation}
Then,
\begin{equation}
	\begin{aligned}
		&\Delta \mathrm{K}-\frac{e^{a/r}\left[r\left(e^{a/r}-3 m\right)+am\right]}{\left(r e^{a/r}-2 m\right)^2}\Delta \mathrm{E}\\
		&-\frac{r^2 e^{a/r}}{\left(r e^{a/r}-2 m\right)}\frac{\partial }{\partial r}\Delta \mathrm{E}=0.
	\end{aligned}
\end{equation}
Therefore, we can obtain a gauge invariant quantity $\psi$
\begin{equation}
	\begin{aligned}
		\psi=&\frac{1}{2} \mathrm{K}-\frac{e^{a / r}\left[a m+r\left(e^{a / r}-3 m\right)\right]}{2\left(e^{a / r} r-2 m\right)^{2}} \mathrm{E}\\
			&-\frac{e^{a / r} r^{2}}{2\left(e^{a / r} r-2 m\right)} \frac{\partial}{\partial r} \mathrm{E}.
	\end{aligned}
\end{equation}
We can employ the same operation to obtain the rest of the gauge invariant quantities
\begin{equation}
	\begin{aligned}
		\delta=&\mathrm{D}+\frac{ e^{a/r}r^2}{2 \left(r e^{a/r}-2 m\right)}\frac{\partial ^2}{\partial t^2}\mathrm{E},\\
		\epsilon=&-\frac{\left[m (a-r) e^{-a/r}\right]}{2 r^2}\mathrm{E}-\frac{\mathrm{A}}{2},\\
		\psi=&\frac{1}{2}  \mathrm{K}-\frac{e^{a / r}\left[a m+r\left(e^{a / r}-3 m\right)\right]}{2\left(e^{a / r} r-2 m\right)^{2}}\mathrm{E}\\
		&-\frac{e^{a / r} r^{2}}{2\left(e^{a / r} r-2 m\right)} \frac{\partial}{\partial r} \mathrm{E},\\
		o=&\frac{\partial}{\partial t}\delta-\frac{\partial}{\partial r}\epsilon\\
		=&\frac{1}{2}\frac{\partial }{\partial r}\mathrm{A}+\frac{\partial }{\partial t}\mathrm{D}+\frac{\left[m \left(a^2-3 ar+r^2\right) e^{-a/r}\right]}{2 r^4}\mathrm{E}\\
		&-\frac{m (a-r) e^{-a/r}}{2 r^2}\frac{\partial }{\partial r}\mathrm{E}+\frac{r^2 e^{a/r}}{2 \left(r e^{a/r}-2 m\right)}\frac{\partial ^2}{\partial t^2}\mathrm{E}.
	\end{aligned}
\end{equation}
\subsection{To solve the gauge invariants from stress-energy tensor}
This part is going to show how we can obtain the solutions of gauge invariants $\psi$ and $o$ as well as deducing the Eq. (\ref{eq38}) from Eqs. (\ref{eq34}) and (\ref{eq37}).

\begin{figure}[htbp]
	\centering
	\includegraphics[width=1.0\linewidth]{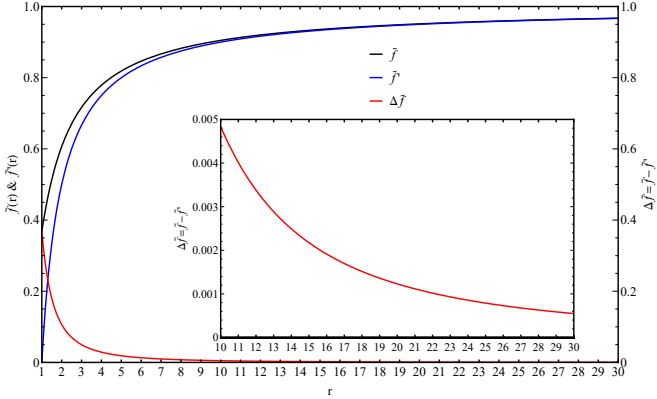}
	\caption{The exact function $\tilde{f}$, the approximation function $\tilde{f'}$ and the residual $\Delta\tilde{f}$ change with the radius $r$. We plot the radius $r$ from the wormhole mouth radius $a$ to perturbation source orbit radius $30a$ as we fix the Misner-Sharp quasi-local mass $m=1.0$, and the unit length of the abscissa axis in the figure is $a$. We use the black solid line to show the exact function $\tilde{f}$, the blue solid line to express the approximation function $\tilde{f'}$ and the red solid line to signify the residual $\Delta\tilde{f}$.}
	\label{appd:1}
\end{figure}

From the stress-energy tensor equation (\ref{eq36}), we can obtain two none-zero components:
\begin{equation}
	\begin{aligned}
		T_{tt}&=M_{mp}\frac{E_{mp}}{R^2},\\
		T_{\phi\phi}&=M_{mp}\frac{L_{mp}}{R^2}.
	\end{aligned}
\end{equation}
And, for any perturbing matter present in the physical spacetime, it can generate a stress-energy source for the linearised Einstein operator (\ref{eq26}). Then, we can obtain the spherical harmonic projection components of linearised Einstein operator. For $E'_{A}$ component:
\begin{equation}
	\begin{aligned}
		E'_{A}&=-16\pi\left(1-\frac{2 m e^{-a / r}}{r}\right)^{2} \oint v^{a} v^{b} T_{a b} Y_{\ell' m'}^{*} \mathrm{d} \Omega\\
		&=-16\pi\left(1-\frac{2 m e^{-a / r}}{r}\right)^{2} \oint \left(1-\frac{2 m e^{-a / r}}{r}\right)^{-1} T_{tt} Y_{00}^{*} \mathrm{d} \Omega\\
		&=-16\pi\left(1-\frac{2 m e^{-a / r}}{r}\right) \oint T_{tt} Y_{00}^{*} \mathrm{d} \Omega\\
		&=-16 \pi\left(1-\frac{2 me^{-a/R}}{R}\right) \frac{M_{mp} E_{mp}}{R^{2}} \delta(r-R) Y_{00}^{*}\left(\theta, \phi\right).
	\end{aligned}
\end{equation}
For $E'_{D}$ component:
\begin{equation}
	\begin{aligned}
		E'_{D}&=16\pi\oint v^{a} Y_{R}^{b *} T_{a b} \mathrm{d} \Omega\\
		&=16\pi\oint v^{a} n^{b} Y_{\ell' m'}^{*} T_{a b} \mathrm{d} \Omega\\
		&=16\pi\oint \left(v^{t} n^{t}T_{tt}+v^{\phi} n^{\phi}T_{\phi\phi}\right) Y_{\ell' m'}^{*}  \mathrm{d} \Omega\\
		&=0.
	\end{aligned}
\end{equation}
For $E'_{E}$ component:
\begin{equation}
	\begin{aligned}
		E'_{E}&=-\frac{16\pi}{2} \oint T_{T 0}^{a b *} T_{a b} \mathrm{d} \Omega\\
		&=-8\pi\oint \left(T_{T 0}^{tt*} T_{tt}+T_{T 0}^{\phi\phi*} T_{\phi\phi}\right)\mathrm{d} \Omega\\
		&=-8 \pi\left(\frac{R-2 me^{-a/R}}{R^3}\right) \frac{M_{mp} L_{mp}}{R^{2}} \delta(r-R) Y_{00}^{*}\left(\theta, \phi\right)\\
		&=-8 \pi\left(1-\frac{2 me^{-a/R}}{R}\right) \frac{M_{mp} L_{mp}}{R^{4}} \delta(r-R) Y_{00}^{*}\left(\theta, \phi\right).
	\end{aligned}
\end{equation}
For $E'_{K}$ component:
\begin{equation}
	\begin{aligned}
		E'_{K}&=-16\pi\left(1-\frac{2 m e^{-a / r}}{r}\right)^{-2} \oint T_{L 0}^{a b *} T_{a b} \mathrm{d} \Omega\\
		&=-16\pi\left(1-\frac{2 m e^{-a / r}}{r}\right)^{-2} \oint n^{a} n^{b} T_{a b} Y_{\ell' m'}^{*} \mathrm{d} \Omega\\
		&=-16\pi\left(1-\frac{2 m e^{-a / r}}{r}\right)^{-2} \oint n^{r} n^{r} T_{r r} Y_{0 0}^{*} \mathrm{d} \Omega\\
		&=0.
	\end{aligned}
\end{equation}
Then, we can combine the two types of linearised Einstein operators as well as Eqs. (\ref{eq34}) and (\ref{eq37})
\begin{equation}
		E_{A}=E'_{A},E_{D}=E'_{D},E_{E}=E'_{E},E_{K}=E'_{K}.
\end{equation}
From the part $E_{D}=E'_{D}$
\begin{equation}
	\frac{4\left(r-2me^{-a/r}\right)}{r^2}\frac{\partial}{\partial t}\psi=0,
\end{equation}
one can find that $\psi\propto t$. From the part $E_{K}=E'_{K}$
\begin{equation}
\frac{4}{r^2}\psi+\frac{4e^{a/r}}{e^{a/r}r-2m}o=0,
\end{equation}
we can obtain that
\begin{equation}
	o=-\frac{e^{a/r}r-2m}{e^{a/r}r^2}\psi,\label{eq103}
\end{equation}
and $o\propto t$. As $\psi\propto t$, the second derivative $\frac{\partial^2}{\partial t^2}\psi=0$. Therefore, the equation $E_{A}=E'_{A}$ and the equation $E_{E}=E'_{E}$ are equivalent. Let's take the equation $E_{A}=E'_{A}$ for solving. As we have mentioned that the perturbation vanishes within the trajectory and our perturbation star's orbit radius is $30r_{H}$, the exponential function $\tilde{f}=e^{-a/r}$ can be expanded in powers as $\tilde{f}=e^{-a/r}=1-a/r+O(r^2)$. We take the function $1-a/r$ as the symbol $\tilde{f'}$, and draw a schematic for the functions $\tilde{f}$ and $\tilde{f'}$ and the residual $\Delta\tilde{f}=\tilde{f}-\tilde{f'}$ changing with the radius $r$ as we fix the Misner-Sharp quasi-local mass $m=1.0$. We can find from the FIG. \ref{appd:1} that the approximation we adopted is valid under the perturbation scene we have chosen.

Therefore, we can rewrite the equation $E_{A}=E'_{A}$ to
\begin{equation}
	\begin{aligned}
		E_{A}=&-\frac{4\left(r-2me^{-a/r}\right)^3}{r^4}\frac{\partial}{\partial r}\psi\\
		&-\frac{4\left[r^2-2m(a-r)\right]\left(r-2me^{-a/r}\right)^2}{r^6}\psi\\
		=&-\frac{4 \left[ 2 a m+r\left(r-2 m\right) \right] ^3}{r^7}\frac{\partial}{\partial r}\psi\\
		&-\frac{4\left[ 2 a m+r\left( r-2 m\right) \right]^2 \left[ r\left(2 m+r\right)-2am\right] }{r^8}\psi\\
		=&-16 \pi\left(1-\frac{2 me^{-a/R}}{R}\right) \frac{M_{mp} E_{mp}}{R^{2}} \delta(r-R) Y_{00}^{*}\left(\theta, \phi\right).
	\end{aligned}
\end{equation}
Next, we can obtain a more obvious expression
\begin{equation}
	\begin{aligned}
		&\frac{\partial}{\partial r}\psi+\frac{\left[ r\left(2 m+r\right)-2am\right] }{r\left[ 2 a m+r\left(r-2 m\right) \right]}\psi\\
		=&-\frac{r^7}{4 \left[ 2 a m+r\left(r-2 m\right) \right] ^3}E'_{A}.
	\end{aligned}
\end{equation}
Then, we can solve this the way we normally solve a linear first-order differential equation
\begin{equation}
	\begin{aligned}
		\psi=C e^{-\int P(r) d r}+e^{-\int P(r) d r} \int Q(r) e^{\int P(r) d r} d r,
	\end{aligned}
\end{equation}
where $C$ is the constant related to the initial conditions, the function $P(r)$ equals to
\begin{equation}
	P(r)=\frac{\left[ r\left(2 m+r\right)-2am\right] }{r\left[ 2 a m+r\left(r-2 m\right) \right]},
\end{equation}
and the function $Q(r)$ equals to
\begin{equation}
	Q(r)=-\frac{r^7}{4 \left[ 2 a m+r\left(r-2 m\right) \right] ^3}E'_{A}.
\end{equation}
Therefore, one can obtain the final solution of $E_{A}=E'_{A}$
\begin{equation}
	\begin{aligned}
		\psi=&2 \sqrt{\pi }\frac{r \eta(r)}{2 a m+r (r-2 m)}\\
		&\times\left\lbrace \frac{M_{mp}E_{mp}R^2}{\left[2 a m+R (R-2 m)\right]\eta(R)}\Theta\left(r-R\right)\right\rbrace.
	\end{aligned}
\end{equation}
Moreover, we should note that $\psi\propto t$. From the formula (\ref{eq103}), we can get the expression of parameter $o$
\begin{equation}
	\begin{aligned}
	   o&=-\frac{e^{a/r}r-2m}{e^{a/r}r^2}\psi\\
	   &=-\frac{a+r-2m}{r (a+r)}\psi\\
	   &=-2 \sqrt{\pi }\frac{\left(a+r-2 m\right)\eta(r)}{(a+r) \left[2 a m+r (r-2 m)\right]}\\
	   &\times\left\lbrace \frac{M_{mp}E_{mp}R^2}{\left[2 a m+R (R-2 m)\right]\eta(R)}\Theta\left(r-R\right)\right\rbrace,
	\end{aligned}
\end{equation}
and the parameter $o$ is also proportional to $t$.

\begin{acknowledgments}
We are grateful to DeChang Dai, Peng Wang and Bo Ning for useful discussions. Wei Hong is grateful to the staffs of the Center for Physical Cosmology of Astronomy Department of Beijing Normal University for warm hospitality. This work is supported by the National Key R$\&$D Program of China (2017YFA0402600), and National Science Foundation of China (Grants No. 11929301, 11573006, 11947408, 12047573).
\end{acknowledgments}

\end{document}